\documentclass[twocolumn]{IEEEtran}
\usepackage{epsfig,amssymb}

\def\BibTeX{{\rm B\kern-.05em{\sc i\kern-.025em b}\kern-.08em
    T\kern-.1667em\lower.7ex\hbox{E}\kern-.125emX}}

\begin{document}
\title{First-Principles Analysis of Molecular Conduction Using Quantum Chemistry Software}
\author{Prashant Damle, Avik W. Ghosh and Supriyo Datta\thanks{datta@purdue.edu}\\
School of Electrical and Computer Engineering, Purdue
University, W. Lafayette, IN 47907}%

\maketitle

\begin{abstract}
We present a rigorous and computationally efficient method to do a
parameter-free analysis of molecular wires connected to contacts. The
self-consistent field approach is coupled with Non-equilibrium Green's
Function (NEGF)
 formalism to describe electronic transport under an applied bias.
Standard quantum chemistry software is used to calculate
 the self-consistent field using density functional theory (DFT). Such
close coupling to standard quantum chemistry software not only makes
the procedure simple to implement but also makes the relation between
the I-V characteristics and the chemistry of
 the molecule more obvious.  We use our method to interpolate between
two extreme examples of transport through a molecular wire connected to
gold (111) contacts: band conduction in a metallic (gold) nanowire, and
resonant conduction through broadened, quasidiscrete levels of a phenyl
dithiol molecule. We obtain several quantities of interest like I-V
characteristic, electron density and voltage drop along the molecule.
\end{abstract}

\section{Introduction}
\label{intro}


There is much current interest 
in molecular electronics
due to the recent success in measuring the I-V
characteristics of individual or small groups of molecules
\cite{reed_review,eigler,collier,chen,reed_expt,zhou_reed_expt,andres_expt,datta_expt}.
The measured resistances exhibit a wide range of values. For example, n-alkane chains
($CH_3-(CH_2)_{n-1}$) have large gaps ($6~eV$ or greater) between
the highest occupied molecular orbital (HOMO) and the lowest
unoccupied molecular orbital (LUMO) and act like strong insulators
\cite{durig,boulas} while gold nanowires (or quantum point contacts)
have zero gap (or a continuous density of states near the Fermi energy)
and exhibit novel one-dimensional metallic conduction characteristics
\cite{ohnishi,yanson}. Similar one-dimensional metallic conduction
(but on a much larger length scale of the order of micrometers)
has been observed in carbon nanotubes \cite{dekker_phys_today}. Then
there are biological molecules like the DNA \cite{fink_dna,porath_dna}
whose electrical conduction characteristics are currently the subject
of much debate \cite{ratner_dna}.  Interesting device functionalities
such as transistor action have also been recently reported \cite{schon_samfet}.

Understanding the correlation between the chemical
and electronic properties of a wide class
of molecules is a first step towards developing a
`bottom-up' molecule-based technology. Semi-empirical theories
\cite{tian,magnus_paper,emberly_prb_58,hush,yaliraki,mujica_vdrop,tsu_theory,magoga_theory,onipko_theory}
have been used to qualitatively study electronic transport in
molecules. Some first-principles methods have also been developed \cite{diVentra,lang_vdrop,guo,seminario,palacios}. Typically 
the first-principles methods 
are either computationally very expensive or
ignore charging effects by employing non self-consistent calculations
. Furthermore, the correct geometry
and atomicity of the contacts is often ignored by employing a jellium-like model although surface
effects like bonding and chemisorption are clearly very important. The
purpose of this paper is to describe a straightforward (computationally
inexpensive) yet rigorous and self-consistent procedure for calculating
transport characteristics while taking into account the effects mentioned
above. In contrast to an ab-initio theory simulating the large (in
principle infinite) open system, we partition the problem into a `device' 
and a `contact' subspace such that standard quantum chemistry techniques
can be employed to analyze the electronic structure of a finite-sized device subspace, 
while incorporating the effects of the outside world (`contacts') through 
appropriately defined self-energy matrices and fields. 

\begin{figure}[t!]
\centerline{\psfig{figure=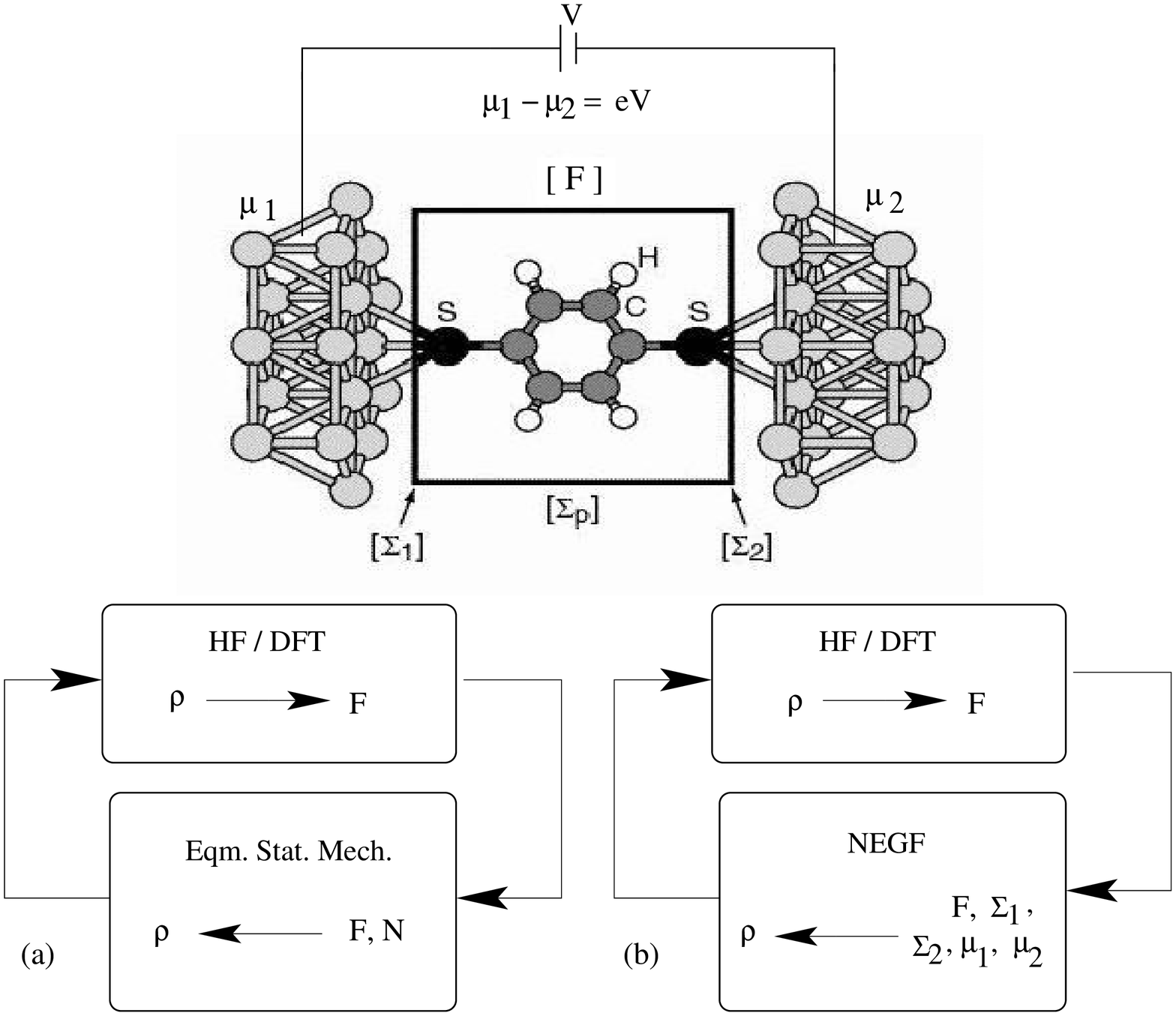,width=3.5in}}
\caption[Schematic of a molecule (phenyl dithiol) connected to two semi-infinite gold contacts]{
Schematic of a molecule (phenyl
dithiol) connected to two semi-infinite gold contacts. The self-energy matrices
$\Sigma_1$ and $\Sigma_2$ {\em exactly} account for the contacts and are of the same size as the 
molecular Fock matrix $F$. The
self-consistency scheme is shown for (a) isolated molecule in equilibrium and (b) contact-molecule-
contact system under bias.}
\label{scheme}
\end{figure}

The electronic structure of an isolated device is obtained in standard
quantum chemistry through a self-consistent procedure \cite{szabo,jensen} 
shown schematically in 
Fig~\ref{scheme}~a. The process consists of two steps
that have to be iterated to obtain a self-consistent solution. Step 1:
Calculate the Fock matrix $F$ for a given density matrix $\rho$ using a specific scheme such as Hartree-Fock (HF) or density 
functional theory
(DFT) to obtain the self-consistent field
, and Step 2: Calculate the density matrix from a given
self-consistent Fock matrix $F$ and total number
of electrons $N$, based on the laws of equilibrium statistical
mechanics.

The problem of calculating the current-voltage (I-V) characteristics
is different from the above self-consistent procedure in the following
ways : (i) We are dealing with an open system having a continuous
density of states and variable (in principle fractional) number of
electrons, rather than an isolated molecule with discrete levels and
integer number of electrons; (ii) The molecule does not necessarily
remain at equilibrium or even close to equilibrium - two volts applied
across a short molecule is enough to drive it far from equilibrium;
(iii) Surface effects like chemisorption and bonding with the contacts
are expected to play a non-trivial role in transport. To calculate
the density matrix (step 2 above) we therefore need a method based on
non-equilibrium statistical mechanics applicable to such an open system
with a continuous density of states. The Non-equilibrium Green's Function
(NEGF) formalism \cite{datta_book,datta_tut} provides us with such a
method and that is what we use in this paper for step 2. Step 1, however,
remains unchanged and we use exactly the same procedure as in standard
quantum chemistry software. The overall procedure is shown schematically
in Fig~\ref{scheme}~b.

We rigorously partition the contact-molecule-contact system (see
Fig~\ref{scheme}) into a contact subspace and a device subspace. This
makes the molecular chemistry conceptually transparent as well as
computationally tractable. The
contact subspace is treated via a one-time calculation of the
surface Green's function of the contacts including their atomicity and
crystalline symmetry. Different molecules coupled to the same contacts
have different couplings but the contact surface Green's function is
independent of the molecule. Given the surface Green's function and the
contact-molecule coupling we can 
describe the chemisorption and bonding of
the molecule with the infinite contacts through self-energy matrices of finite size (equal to that of the molecular subspace).
The NEGF formalism has clear prescriptions
(Section~\ref{theory}) to calculate the non-equilibrium density matrix
from a knowledge of $F$, $\Sigma_{1,2}$ and the electrochemical
potentials in the two contacts, $\mu_1$ and $\mu_2$. All quantities of
interest (electron density, current etc) are then calculated from the
self-consistently converged density matrix.

From a computational viewpoint, the most challenging part is the
calculation of the Fock matrix $F$, which involves the core molecular 
Hamiltonian and the self-consistent potential. A number of researchers have 
developed their own schemes for
performing such ab-initio computations \cite{diVentra,lang_vdrop,guo}. We accomplish this part 
by  exploiting the fast algorithms of Gaussian '98
\cite{gaussian}, a commercially available quantum chemistry software.
Aside from the computational advantages of using an already well-established
software, such a close coupling to the standard tools for analyzing molecules 
makes the chemistry of the system clearer. 
 We describe the contacts and the molecule using the sophisticated LANL2DZ
basis set \cite{lanl2dz_1,lanl2dz_2} which incorporates relativistic
core pseudopotentials. The self-consistent potential is calculated using
DFT with Becke-3 exchange \cite{becke}
and Perdew-Wang 91 correlation \cite{perdew}. Thus, equipped with a
SUN workstation, we are able to perform a parameter-free analysis of
conduction in molecular wires in a few hours \cite{damghosh}.

We illustrate our method using a gold nanowire and a phenyl dithiol
molecule sandwiched between gold contacts and study some previously
addressed issues like (1)  charge transfer and self-consistent
band lineup (e.g. \cite{xue_paper}), (2) I-V characteristics
(e.g. \cite{tian,emberly_prb_58,diVentra})  and (3)  charge density
and voltage drop (e.g.  \cite{lang_vdrop,mujica_vdrop}). Metallic
conduction with quantum unit conductance is observed in the gold
nanowire. Upon reducing the coupling to contacts, the gold nanowire
exhibits resonant tunneling type of conduction just as seen in a phenyl
dithiol molecule. The introduction of a defect (a stretched bond) in
the nanowire gives rise to a sharp voltage drop across the impurity
as expected. The presence of the defect leads to negative-differential
resistance (NDR) in the I-V characteristic of the wire. Periodic Friedel 
oscillations are observed in the charge density near the defect, the magnitude of 
these oscillations decreasing as expected upon the introduction of phase-breaking 
scattering. 

This paper is organized as follows.
Section~\ref{theory} contains a fairly detailed description of the
theoretical formulation, specifically modeling the influence of the contacts on 
the device subspace through self-energy matrices, and developing an appropriate
transport formalism to calculate the density matrix $\rho$ (Step 2 above) for the resulting open system under bias.
The calculation of the Fock matrix $F$ given the density matrix 
$\rho$ (Step 1 above) is a standard procedure in quantum chemistry and we will not discuss it further. 
Section~\ref{results} shows the results and
Section~\ref{summary} briefly summarizes the paper.

\hfill

\section{Theoretical Formulation}
\label{theory}

\subsection{Broadening in an open system: Self-energy.} 
The concept of self-energy is used in many-body physics to describe non-coherent electron-electron and
electron-phonon interactions. We could do the same in principle and use a self-energy function $\Sigma_p$ to describe the
effect of non-coherent interactions of the molecule with its surroundings (see Fig~\ref{scheme}). 
For the most part of this paper (one exception is made in Fig~\ref{vdrop}~a, bottom panel), we will neglect 
such non-coherent scattering processes ($\Sigma_p=0$) 
because the
experimental current-voltage characteristics do not show any signatures of the molecular vibration spectra.
In addition to $\Sigma_p$,
we can use self-energy functions
$\Sigma_1$ and $\Sigma_2$ to describe the interactions of the molecule with the two contacts respectively
\cite{datta_book}. 

The contact self-energies $\Sigma_1$ and $\Sigma_2$ arise formally out of partitioning an infinite
system and projecting out the contact Hamiltonians.  When an isolated
molecule with discrete energy levels is contacted to leads to make an
infinite composite system, the energy-dependent one-particle retarded
Green's function of the complete system is expressed in an appropriate
basis set as:
\begin{equation}
G(E) = \left[(E+i0^+)S - F \right]^{-1}
\label{e1}
\end{equation}
where $S$ is the overlap matrix and $F$ is the Fock matrix for
the whole system. $F$ incorporates the effect of external fields, the
electronic kinetic energy, electron-nuclear attractions, as well
as electron-electron interactions, which could in principle include
Coulomb, exchange and correlation effects. The poles of this Green's function
lie near the real energy axis, and represent the energy levels for the
infinite system. To extract just the device part of $G$ involving the device
overlap matrix $S_{dd}$ and the device Fock matrix $F_{dd}$, we utilize
the fact that for a matrix
\begin{eqnarray}
G & = & (ES-F)^{-1} \nonumber \\ 
 & = & \left[\begin{array}{ccc} ES_{dd}-F_{dd} &\vdots& \tau \\ \cdots
&&\cdots \\ \tau^\dagger & \vdots &
D\end{array}\right]^{-1} \nonumber \\
 & = & \left[\begin{array}{ccc} G_{dd} &\vdots&   \\ \cdots
&&\cdots \\  &  \vdots &
 \end{array}\right]
\label{e2}
\end{eqnarray}
the device part $G_{dd}$ is given by $\left(ES_{dd}-F_{dd} -
\tau D^{-1}\tau^\dagger\right)^{-1}$. $\tau D^{-1}\tau^\dagger $ is the self-energy term that describes the effect of the 
contacts on the device. Only a few surface (interface between device and contact)
sites give rise to non-zero coupling elements in $\tau$ \cite{datta_book}, and thus only the surface 
term of $D^{-1}$, the
surface Green's function $g$ is needed. We are thus left with a reduced device
Green's function $G_{dd}$ given by:
\begin{equation}
G_{dd}(E) = \left[ES_{dd} - F_{dd} - \Sigma_1(E) - \Sigma_2(E)\right]^{-1}
\label{e3}
\end{equation}
where the self-energy matrices $\Sigma_1$ and $\Sigma_2$ are {\em non-Hermitian} matrices arising
from partitioning out contacts 1 and 2 respectively.  Each contact self-energy matrix
is related to the non-zero part of the corresponding lead-device coupling $\tau$ and the
surface Green's function $g(E)$ through:
\begin{equation}
\Sigma_{1,2} = \tau_{1,2}g_{1,2}\tau_{1,2}^\dagger
\label{e4}
\end{equation}

The geometry of the bonding between the molecule and the contact surface determines the coupling matrices $\tau_{1,2}$.
For thiol bonds for example, experimentally
it is believed that a chemically bonded sulfur atom on a gold surface
overlaps equally with three gold atoms that form an equilateral triangle
as shown in Fig~\ref{scheme} \cite{larsen,camillone}.  We use the LANL2DZ
basis set \cite{lanl2dz_1,lanl2dz_2} to describe both the contacts and
the molecule. LANL2DZ is a sophisticated basis set with relativistic
core pseudopotentials and is observed to provide a good description of 
the contact surface density of states around the Fermi energy \footnote{For
a proper description of the bandstructure of gold away from the Fermi energy,
we observe that it is necessary to include upto the 4th or 5th nearest
neighboring interactions, because some of the basis functions in LANL2DZ
are relatively delocalized.}.  The coupling
matrices $\tau_{1,2}$ are calculated by using Gaussian '98 to simulate an
`extended molecule' \cite{xue_paper,xue_thesis} consisting of the molecule
under consideration and an equilateral triangle of three gold atoms
on either side of the molecule. 

\begin{figure}[t]
\centerline{\psfig{figure=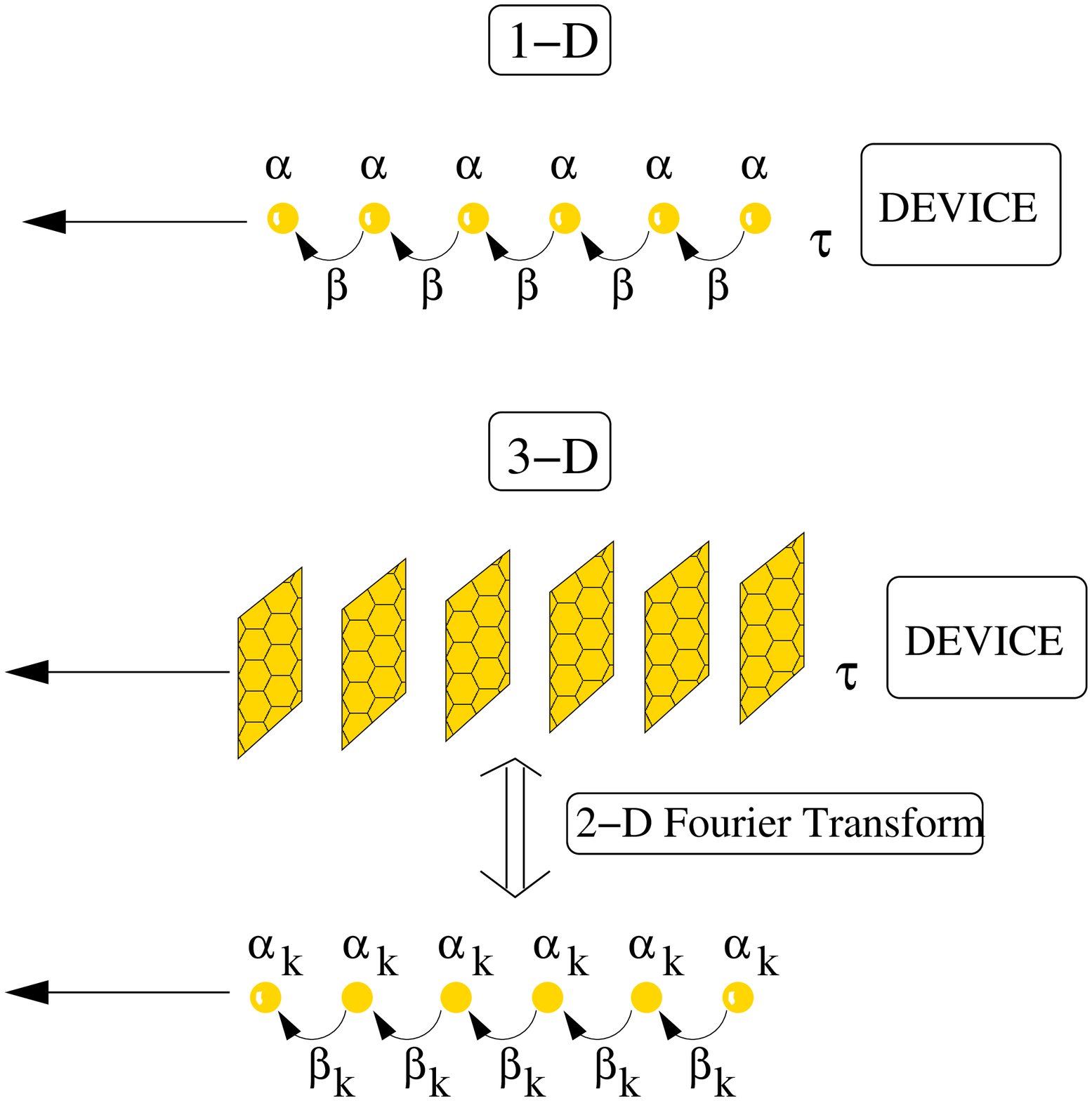,width=3.5in}}
\caption[Device coupled to a semi-infinite contact]{Device coupled to a semi-infinite 
contact (top: 1-D contact, bottom: 3-D
contact). The contact
self-energy depends on the device-contact coupling $\tau$ and the contact surface 
Green's function $g$ which is recursively
solved for by using the contact on-site and coupling matrices $\alpha$ and $\beta$
(see text). In case of a 3-D contact, the
in-plane periodicity is used to obtain an equivalent 1-D picture with k-dependent $
\alpha_k$ and $\beta_k$.}
\label{1dlead}
\end{figure}

The surface Green's function matrices $g_{1,2}$ 
are obtained recursively for a periodic lattice by the same decimation
process as for the device. Removing one layer of the contact lattice
gives back the same surface Green's function, so each contact surface
Green's function satisfies a recursive equation \cite{manoj_thesis} involving on site
and coupling matrices $\alpha$ and $\beta$ (diagonal and 
off-diagonal blocks of $ES-F$ respectively, see Fig~\ref{1dlead}):
\begin{equation}
g^{-1} = \alpha - \beta g\beta^\dagger
\label{e5}
\end{equation}
To generalize to a 3-D
lead of a specific orientation, we follow the procedure in \cite{manoj_thesis} and go to the 2-D $k$-space representation
for each cross-sectional plane of the lead, so that each $k$-point effectively
acts as an independent 1-D problem for which the above recursive formula
holds. 

\begin{figure}[t!]
\centerline{\psfig{figure=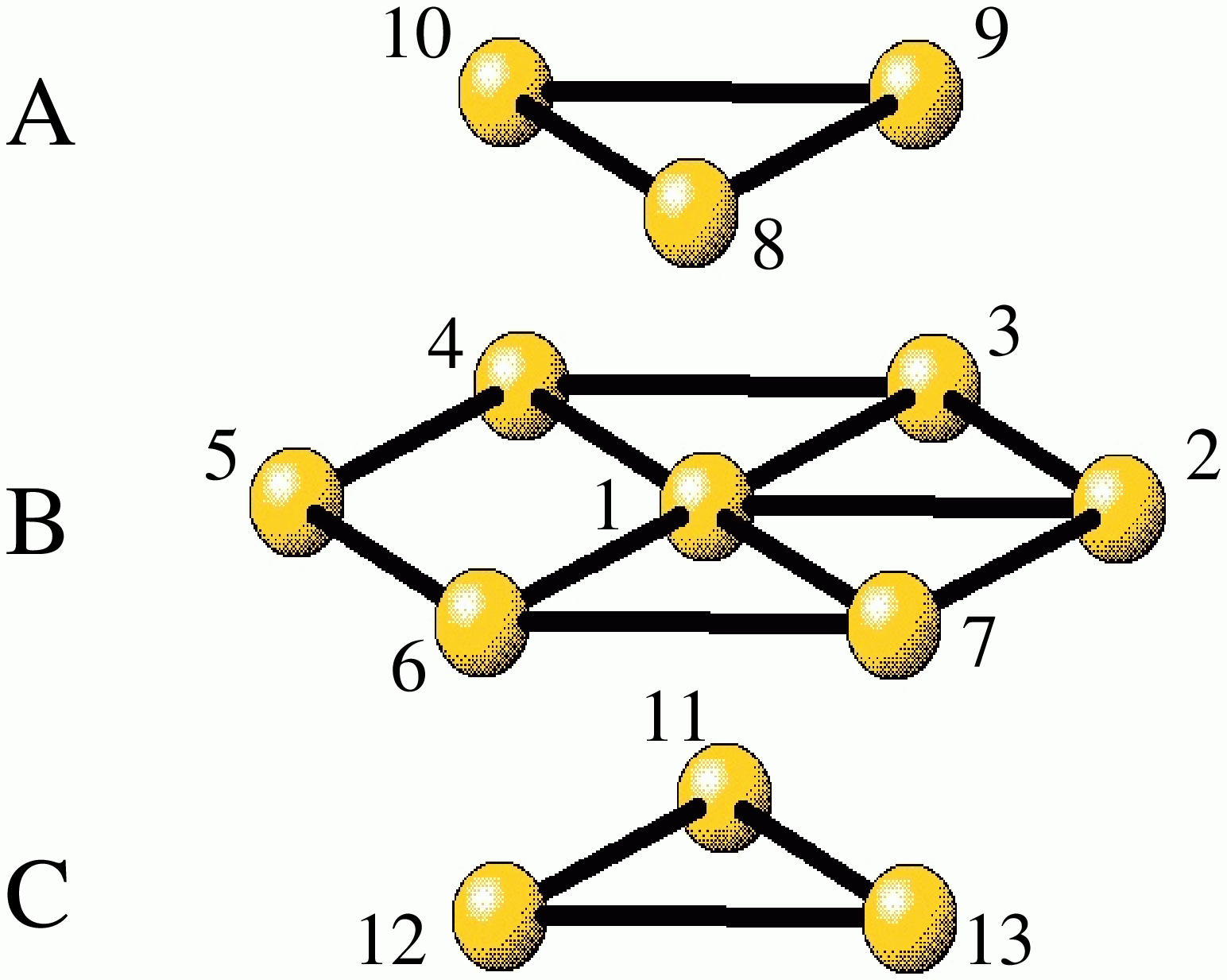,width=3.5in}}
\caption[Central $13$ atom part of the 28 atom gold cluster in FCC (111) geometry]{
Central $13$ atom part of the 28 atom gold
cluster in FCC (111) geometry used to calculate the coupling matrices
$S_{mn}$ and $F_{mn}$ in Eq.~\ref{FakSak}. A reference atom (atom 1) and all it's 
nearest neighbors are shown in this central
13 atom part. 28 atoms are used so as to reduce the `edge' effects that
tend to destroy the symmetries associated with the Fock matrices
$F_{mn}$ (see the discussion following Eq.~\ref{FakSak} and
Appendix~A).}

\label{cluster}
\end{figure}

For gold (111) leads, we use Gaussian '98 to extract in-plane and out-of-plane nearest-neighbor overlap matrix $S$ and Fock
matrix $F$ components using a gold cluster (see Fig~~\ref{cluster}) and define Fourier components in the (111) 
plane:
\begin{eqnarray}
F_{a\vec{k}} = \sum_nF_{mn}e^{-i\vec{k}\cdot(\vec{r}_m - \vec{r}_n)}\nonumber\\
S_{a\vec{k}} = \sum_nS_{mn}e^{-i\vec{k}\cdot(\vec{r}_m - \vec{r}_n)}
\label{FakSak}
\end{eqnarray}
where $m$ is an arbitrary gold atom, and $n$ involves a sum over $m$ and all its nearest in-plane neighbors, with coordinate
$\vec{r}_n$. 
The out-of-plane Fourier components $F_{b\vec{k}}$ and $S_{b\vec{k}}$ are also defined analogously.
The $F_{mn}$ and $S_{mn}$ matrices must all obey the group theoretical symmetry of the FCC crystal. This symmetry 
is satisfied by the $S_{mn}$ matrices since the overlap between two atoms depends only on those two atoms and not on the rest of 
the atoms in the gold cluster. The $F_{mn}$ matrices, however, are obtained via a self-consistent calculation (see 
Fig~\ref{scheme}~a) and depend on the presence or absence of other atoms in the cluster. Ideally we need to simulate 
an infinite 
cluster (or crystal) in order to get the Fock matrices to obey the symmetry. For practical reasons we simulate a finite cluster 
consisting of 28 gold atoms arranged in the FCC (111) geometry (the central 13 atom part of this cluster is shown in 
Fig~\ref{cluster}) and enforce the known symmetry rules for FCC  
(111) crystal structure on the resulting Fock matrices. A 
detailed description of this procedure of enforcing symmetry is given in Appendix~A. With the correct symmetry imposed, the 
$F_{a\vec{k}}$ and $S_{a\vec{k}}$ matrices in Eq.~\ref{FakSak} are Hermitian. The gold surface Green's function in $\vec{k}$ 
space is then obtained by iteratively solving \cite{manoj_thesis} 
\begin{equation}
g^{-1}_{\vec{k}} = \alpha_{\vec{k}} - \beta_{\vec{k}}g_{\vec{k}}\beta^\dagger_{\vec{k}}
\label{gk}
\end{equation}
where
\[
\alpha_{\vec{k}} = (E+i0^+)S_{a\vec{k}} - F_{a\vec{k}} 
\]
and
\[
\beta_{\vec{k}} = (E+i0^+)S_{b\vec{k}} - F_{b\vec{k}}
\]
The real-space gold (111) surface Green's function matrices are then obtained using 
\begin{equation}
g_{mn} = \frac{1}{N}~\sum_{\vec{k}}g_{\vec{k}}e^{i\vec{k}\cdot(\vec{r}_m - \vec{r}_n)}
\label{greal}
\end{equation}
where $N$ is the number of unit cells in the (111) plane (or the number of $k$ points). The surface Green's function matrices so
obtained are independant of the molecule under consideration and depend only on the material and geometry of the contacts.

We have thus managed to partition the system {\it{exactly}} into a
device and a lead subspace. The self-energy matrices replacing the
contacts are non-Hermitian, their real parts representing the shift in
the molecular energy levels due to coupling with the infinite contacts,
and their imaginary parts representing the broadenings of these levels
into a continuous density of states. 
The partitioning makes the problem
computationally tractable, since the size of the self-energy matrices
is the same as that of the device Fock matrix, even though they
represent the effect of infinitely large contacts exactly. In addition, the partitioning
decomposes the problem into three different subspaces each involving
a different area of research: (i) the device Fock matrix $F_{dd}$
incorporates the quantum chemistry of the intrinsic molecule; (ii)
the coupling matrix $\tau$ involves details of the bonding between the
molecule and the contact (chemisorption, physisorption etc.) and (iii)
the surface Green's function $g$ involves the surface physics of the metallic
contact, which could in principle be extended to include additional
effects such as surface states, band-bending, surface adsorption and
surface reconstruction.

It is desirable to include a few metal atoms as part of the device for a number of reasons. (i) The molecule may affect a few
nearby atoms on the surface of the metallic contact. These surface effects will be automatically accounted for in the
self-consistent calculation if a few surface metal atoms are included as part of the device. (ii) Density functional theory is
traditionally used for a finite system with an integer number of electrons, or periodic systems. Extending DFT to a non-neutral
open molecular subsystem with fractional number of electrons is a topic of intense research \cite{yang_frac_el,russier_frac_el}.
However an extended molecule which includes a few metal atoms is effectively charge neutral and allows the standard DFT
formalism to go through.  (iii) The charges in the molecule are imaged on the metallic contact. These image charges need to be
taken into account in order to accurately calculate the self-consistent potential. It is reasonable to expect that the image
charges will reside on a few surface metal atoms that are close to the molecule, and hence the inclusion of these surface atoms
in the device should account for the image charge effect on the self-consistent potential. (iv) Finally, the atoms near the two
ends of the molecule will have slightly erroneous charge densities because the device and contact basis functions are not
orthogonal to each other and the partitioning of charge leads to ambiguities at the interfaces (see Mulliken/L\"owdin
\cite{lowdin} partitioning). So it is desirable to `pad' the molecule with a few metal atoms on the two ends so as to allow an
accurate calculation of the molecular charge.

The results we present in this paper are obtained using just the molecule as our `device' (metal atoms are not included in the
device) in order to reduce the computational time. Preliminary calculations indicate that including the metal atoms improves 
certain aspects such as the charge density on the end atoms. However, such a calculation with an extended device takes much more 
computer time, especially with gold contacts. 

\begin{figure}[t!]
\centerline{\psfig{figure=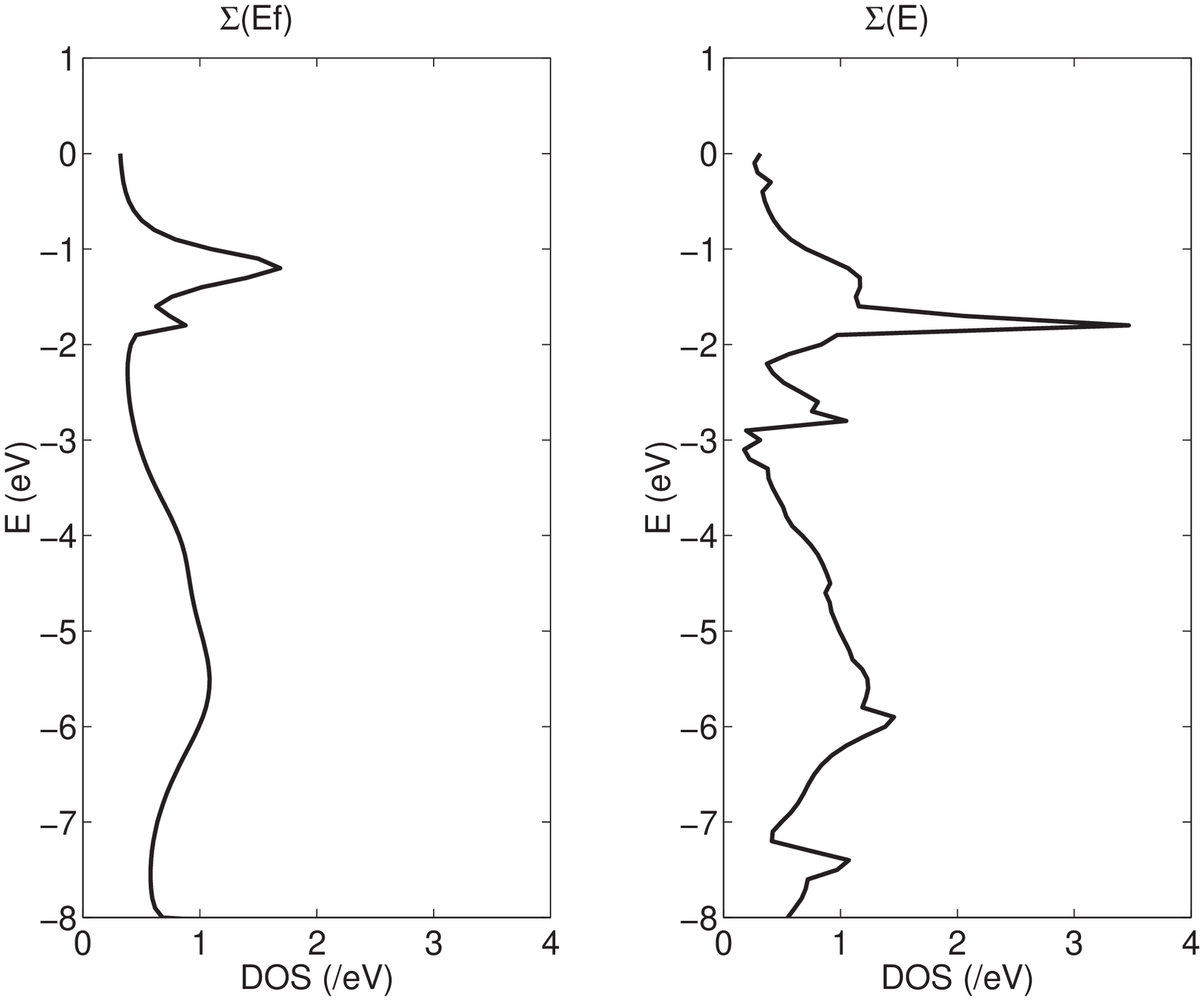,width=3.5in}}
\caption[Comparison of non-self-consistent evaluations of DOS for phenyl dithiol wi
th and
without a constant self-energy approximation]{Comparison of non-self-consistent
evaluations of DOS for phenyl dithiol with and
without a constant self-energy approximation. The excellent correspondence is
a consequence of the relatively flat DOS of Au(111) near the Fermi energy.}
\label{compare}
\end{figure}

The energy dependent self-energy matrices discussed above exactly
account for the bonding and chemisorption of the molecule onto the
contact surface, as we discuss in our results section. For gold (111)
contacts we find that these issues are taken care of even if we replace
$\Sigma(E)$ with $\Sigma(E_f)$ ($E_f$ is the gold Fermi energy) as is
evident from Fig.~\ref{compare}. This helps to reduce the time taken to
compute the density matrix (see Eq.~\ref{rho}). We have developed a
fast and elegant analytical method to evaluate the density matrix for
an energy independent self-energy. This method is explained in
Appendix~B. Such a simplification may not be possible for platinum
contacts having a significant structure in the density of states near
$E_f$. This makes the calculation of the density matrix computationally
quite challenging because the correlation function may have sharp peaks
in the range of integration. Integrating over a complex energy contour
\cite{xue_paper,contour} simplifies the computational complexity to a
certain degree, but the energy range of integration has to be huge in
order to take into account all the molecular levels (see Appendix~B),
and a faster scheme is desirable.

\subsection{Transport: Non-equilibrium Green's function (NEGF) formalism.} 

The NEGF
formalism provides a suitable method for calculating the density matrix
$\rho$ for systems under non-equilibrium conditions. A tutorial
description of this formalism can be found in \cite{datta_tut}. Here we
will simply summarize the basic relations that can be used (1) to obtain
$\rho$, given the molecular Fock matrix $F$, the self-energy matrices
$\Sigma_1$, $\Sigma_2$, and the contact electrochemical potentials
$\mu_1$ and $\mu_2$ and (2) to obtain the electron density $n(\vec{r}$)
and current $I$ from the self-consistent density matrix $\rho$. For
open systems with a continuous density of states, the density matrix
can be expressed as an energy integral over the correlation function
$-iG^<(E)$, which can be viewed as an energy-resolved density matrix:
\begin{equation} 
\rho = \int dE[-iG^<(E)/2\pi] \label{rho} 
\end{equation}
The correlation function is obtained from 
\begin{equation} 
-i{G}^< = G\left({f_1\Gamma_1
 + f_2\Gamma_2}\right)G^\dagger 
\label{e8}
\end{equation} 
where $f_{1,2}(E)$ are the Fermi functions with
electrochemical potentials $\mu_{1,2}$ 
\begin{equation} 
f_{1,2}(E) = \left( 1 + \exp{\left[ {{E-\mu_{1,2}}\over{k_BT}}\right]}\right)^{-1}
\label{e9} 
\end{equation} 
G is the Green's function matrix
(energy-dependent) in a non-orthogonal basis: 
\begin{equation} {G} =
\left(ES - F - {\Sigma}_1 - {\Sigma}_2\right)^{-1} 
\label{e10}
\end{equation} 
where $S$ is the overlap matrix. 
The broadening functions
are the anti-Hermitian components of the self-energy: 
\begin{equation}
{\Gamma}_{1,2} = i[{\Sigma}_{1,2} - {\Sigma}_{1,2}^\dagger] 
\label{e11}
\end{equation}
The NEGF equation for $\rho$ can be interpreted as the filling of broadened energy levels 
(eigenvalues of $F$) by two separate contacts, with $G\Gamma_i G^\dagger$ representing the
density of states (DOS) contribution from the $i$th electrode, and $f_i$ representing the Fermi function of 
the $i$th electrode. Eq.~\ref{e8} assumes that the contacts are reflectionless \cite{datta_book}.

The converged (see Fig~\ref{scheme}~b) density matrix is used to obtain the 
total number as well as the spatial distribution of electrons using 
\begin{eqnarray}
N &=& {\rm trace} (\rho S)\nonumber\\
n(\vec{r}) &=& \sum_{m,n}\rho_{mn}\phi_m(\vec{r})\phi_n(\vec{r})
\label{e12}
\end{eqnarray}
where $\phi_{m,n}(\vec{r})$  represent molecular basis functions. The 
converged density matrix may also be used to obtain the terminal current \cite{datta_book}.
For coherent transport
\footnote{
We can modify the transmission formalism to model incoherent scattering via B\"uttiker probes \cite{buttiker}.
We associate a dephasing strength $\eta$ and a single electrochemical potential $\mu_p$ with each 
B\"uttiker probe. $\mu_p$ is obtained iteratively by asserting that the {\em total} current injected by the B\"uttiker probes 
is zero ($f_p$ is the Fermi function with electrochemical potential $\mu_p$):
\[
(2e/h)\left[\int dE~T_{1p}~\left(f_p-f_1 \right) + \int dE~T_{2p}~\left(f_p-f_2 \right) \right] = 0
\]
This implies that we model incoherent {\em inelastic} scattering as opposed to incoherent {\em elastic} scattering which is 
commonly used \cite{tian,li_dna} to model phase breaking processes. This procedure was used for the bottom panel in 
Fig.~\ref{vdrop}~a.}
, we can simplify the calculation of the current by using
the transmission formalism where the transmission function 
\cite{datta_book}:
\[
T(E) = {\rm trace} \left[ \Gamma_1 G \Gamma_2 G^\dagger \right]
\]
is used to calculate the terminal current
\begin{equation}
I = (2e/h)\int_{-\infty}^\infty dE~T(E)~ 
\left(f_1(E)-f_2(E) \right)
\label{e13}
\end{equation}

\section{Results}
\label{results}

\subsection{Equilibrium}

\begin{figure}[t!]
\centerline{\psfig{figure=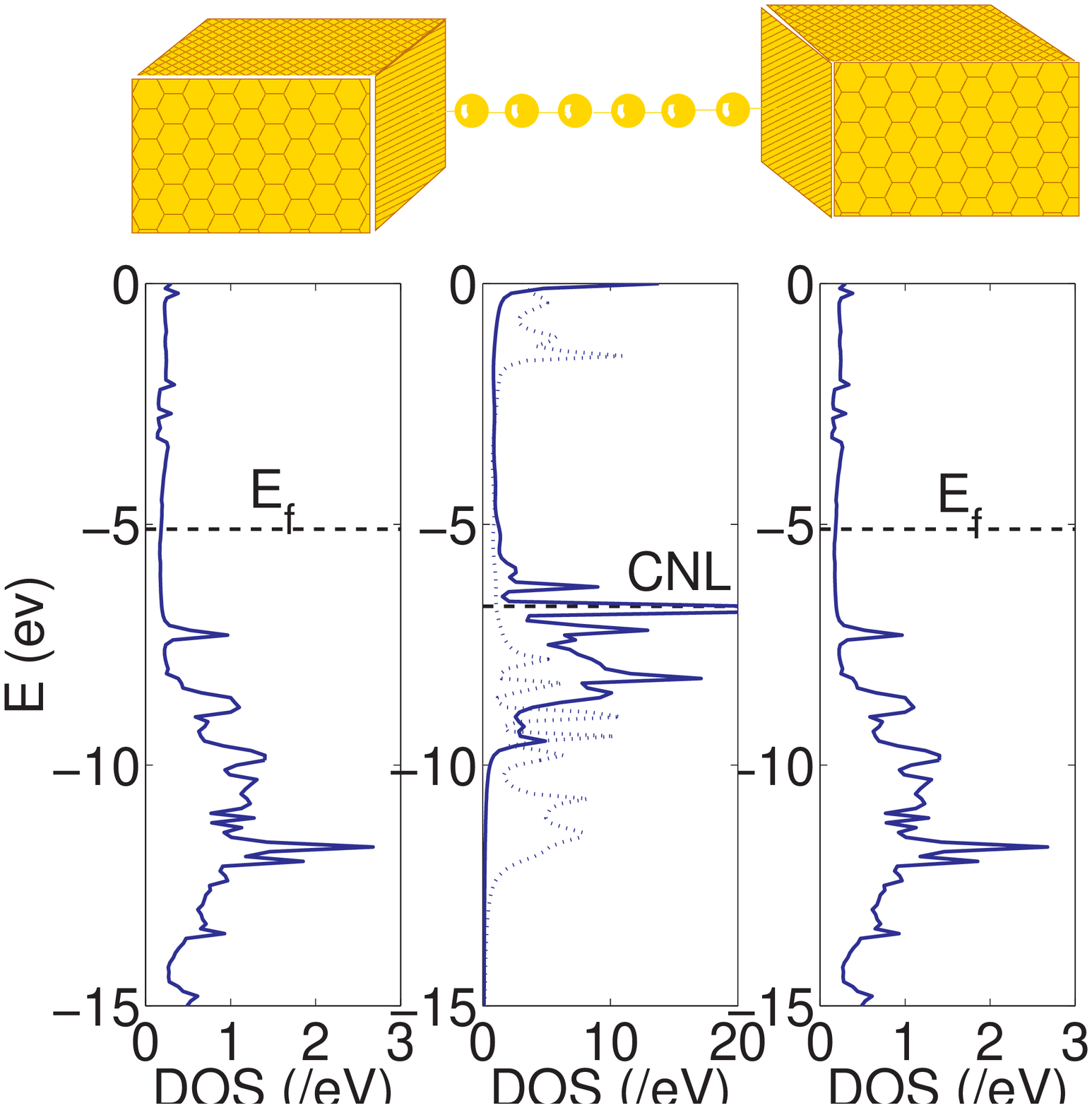,width=3.5in}}
\caption[Charge transfer and self-consistent band lineup upon coupling a gold
nanowire to two gold contacts]{ Charge transfer and self-consistent band lineup
upon coupling a gold
nanowire to two gold contacts. The charge neutrality level (CNL) of the wire
lies below the Fermi energy of gold, leading to electron flow from
contacts to the wire. The DOS (dashed) of the charge-neutral wire redistributes and
 floats up (solid)
self-consistently due to charging energies associated with the electron flow.}
\label{cnl}
\end{figure}

{\em Charge transfer and band lineup.} The charge transferred
between a device and contacts leads to aligning of reference energy levels
in the two subsystems. Such band lineup issues are of primary importance
in understanding semiconductor heterojunctions as well as plastic electronics. 
An analogous band-lineup diagram can be obtained on
coupling a molecule to contacts. 
Fig~\ref{cnl} shows the surface DOS of the contacts and the DOS
of the device (a gold nanowire in this case). The 6 atom gold chain has
discrete energy levels that are broadened into a metallic band-like
DOS on coupling to contacts. For the chain with a broadened density of
states, one can define a charge-neutrality level (CNL) \cite{tersoff},
such that filling up all the states below the CNL
keeps the device charge-neutral. The Fermi energy of gold, approximately
-5.1 eV (negative of the bulk gold work function), is noticeably
higher than the CNL of a 1-D gold nanowire. Owing to this difference
electrons flow in from the contacts to the device.  For a small molecule,
the capacitative charging energy is large, so the original broadened levels
(dashed) float up self-consistently (solid) due to charging till the charge
transfer ceases, the band lineup is complete and the device and contacts are
in chemical equilibrium. 

\subsection{Non-equilibrium}

\begin{figure}[h!]
\centerline{\psfig{figure=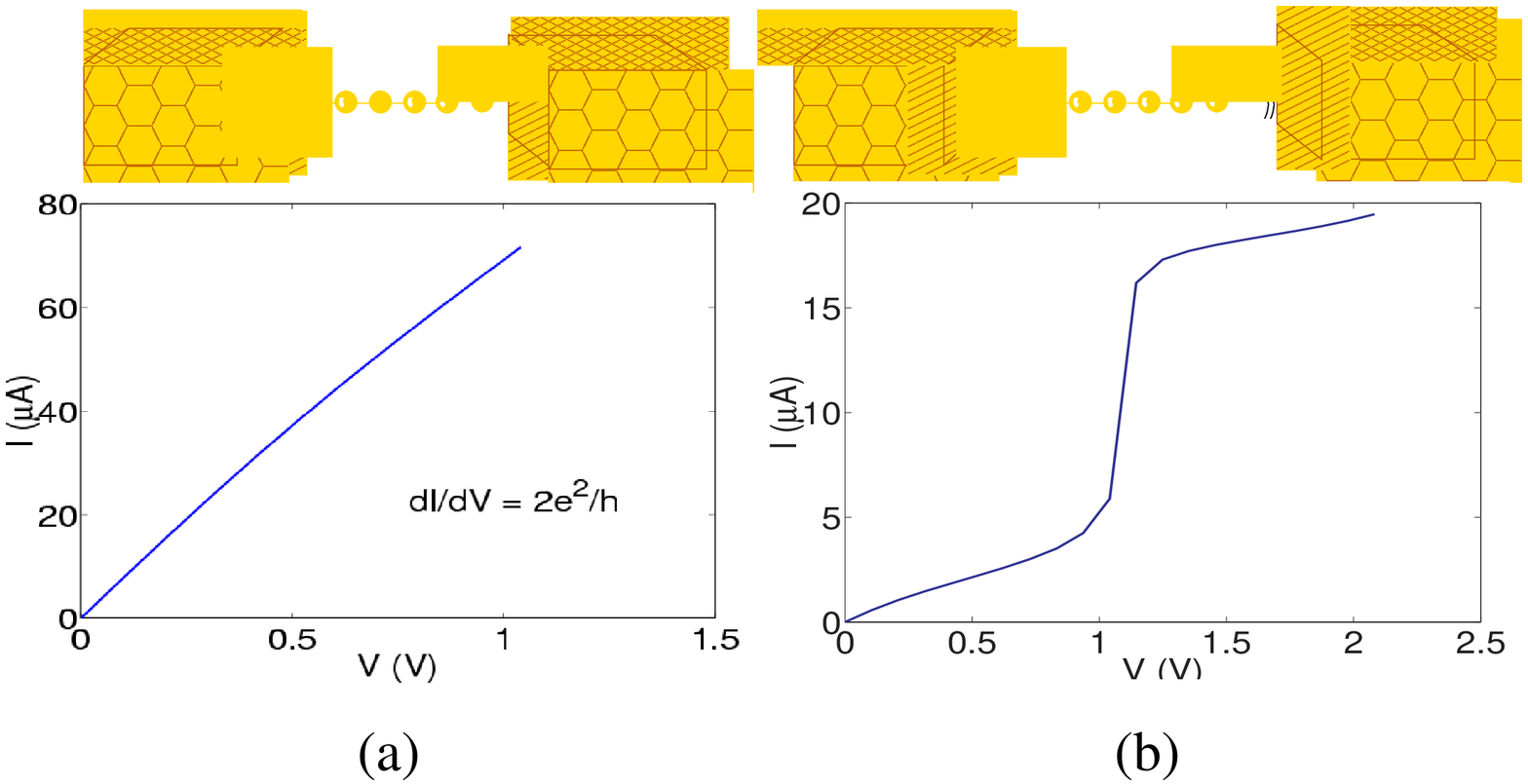,width=3.5in}}
\caption[I-V of a gold nanowire coupled to gold contacts]{(a)I-V of a Gold nanowire
 with six gold atoms forming a 1-D chain
connected to two gold contacts. An
energy-independent self-energy $\Sigma(E_f)$ is used to model the contacts. A quantum 
unit conductance of $2e^2/h$ signifies
that $\Sigma(E_f)$ accounts for the perfect transmission by seamlessly joining the
nanowire to the contacts. (b) I-V of the
same gold nanowire but with reduced contact coupling which decreases level broadening 
and a resonant tunneling type conduction is
seen. The I-V shows step-like behavior with the current jumping up when the contact
 Fermi levels $\mu_1$ and/or $\mu_2$ cross a
molecular level (See Figs~\ref{lev_cartoon}, \ref{lev_weak}).}
\label{iv_qpc}
\end{figure}

\begin{figure}[h!]
\centerline{\psfig{figure=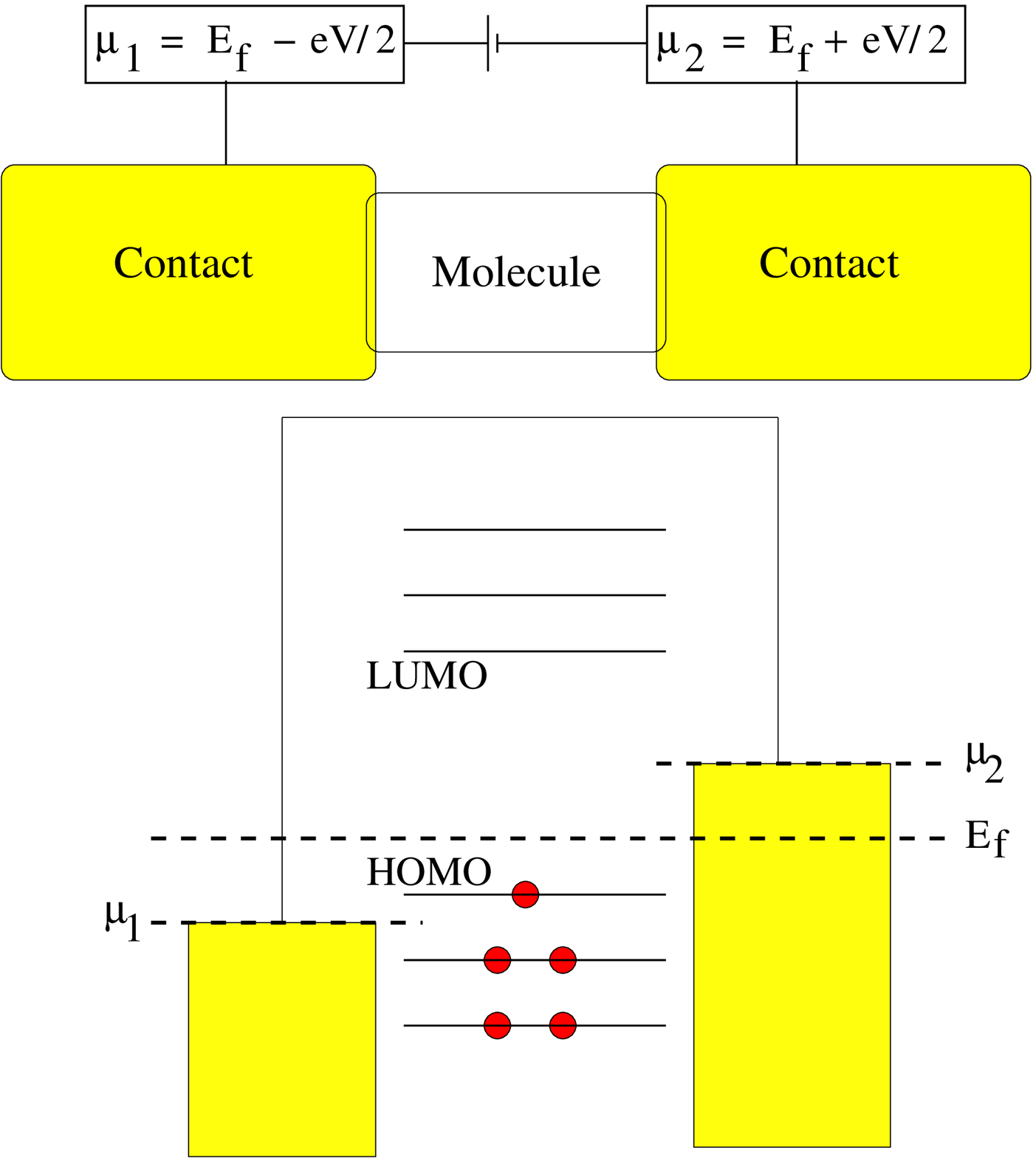,width=3.5in}}
\caption[mechanism]{Mechanism explaining how current flows through a molecule coupled to contacts.
At equilibrium ($V=0$) $\mu_1=\mu_2=E_f$ and no
current flows. For non-zero $V$ the contact Fermi energies separate by
an amount $eV$ 
and a significant current flows only when a
molecular level lies in between $\mu_1$ and $\mu_2$ (also see Fig~\ref{lev_weak}).}
\label{lev_cartoon}
\end{figure}

\begin{figure}[h!]
\centerline{\psfig{figure=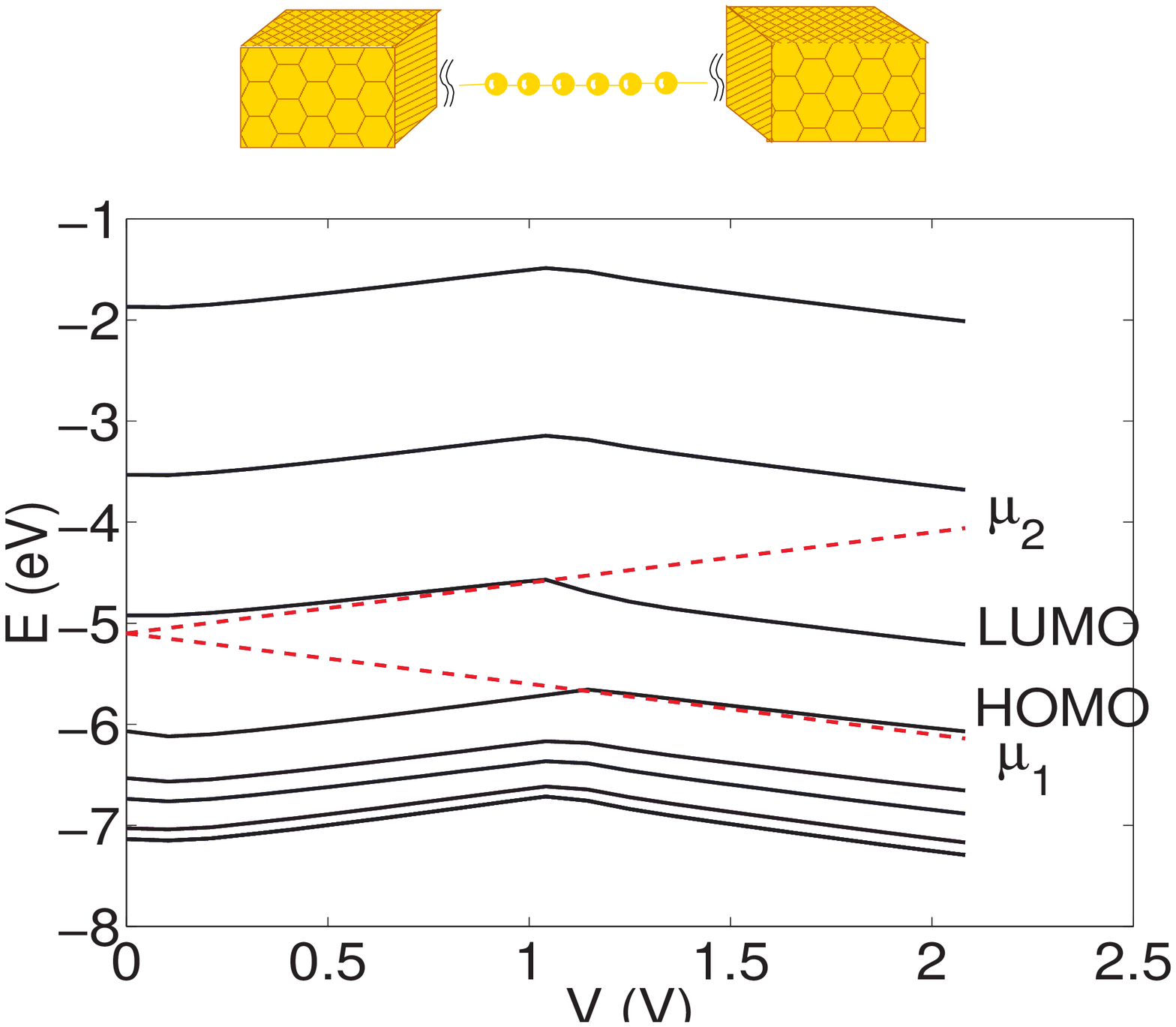,width=3.5in}}
\caption[Energy levels of a gold nanowire weakly coupled to contacts as a function
of applied bias]{Energy levels of a gold
nanowire weakly coupled to contacts as a function of applied bias. The I-V for this
 case is
shown in Fig~\ref{iv_qpc}~b. At zero bias the contact Fermi level is close to LUMO.
 At non-zero bias $\mu_2$ wants to fill
up the LUMO and add extra electrons in the device, so charging effects tend to float 
the levels up till $\mu_1$
comes close to HOMO and wants to empty it. At this point, more electrons are lost 
than gained and charging effects
make the levels float down parallel to $\mu_1$ in order to minimize the loss of 
electrons.}
\label{lev_weak}
\end{figure}

\begin{figure}[h!]
\centerline{\psfig{figure=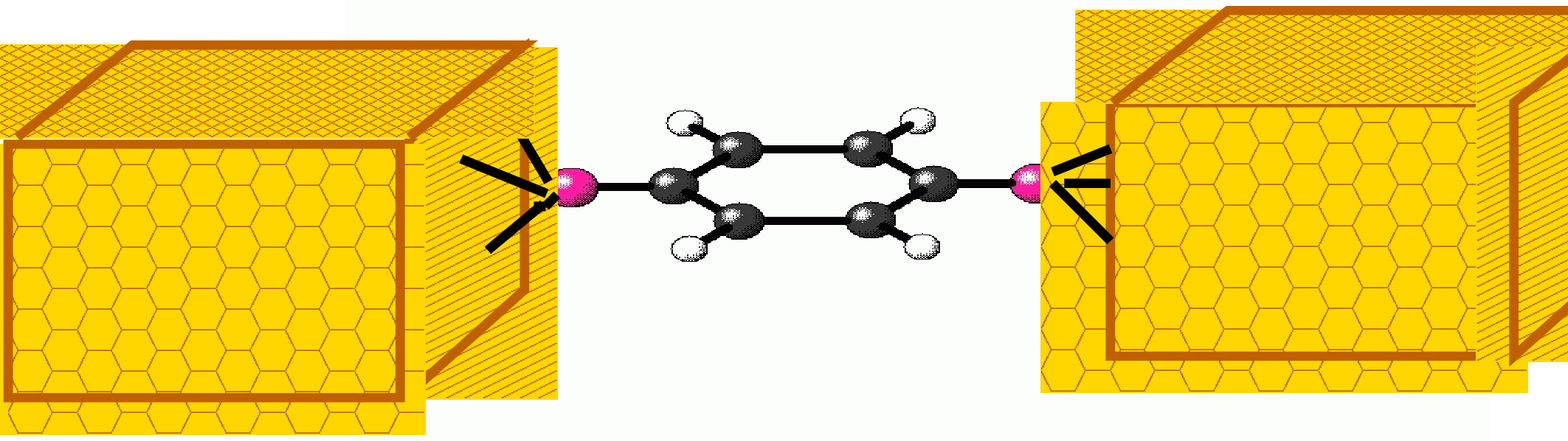,width=3.5in}}
\vspace{0.2cm}
\centerline{\psfig{figure=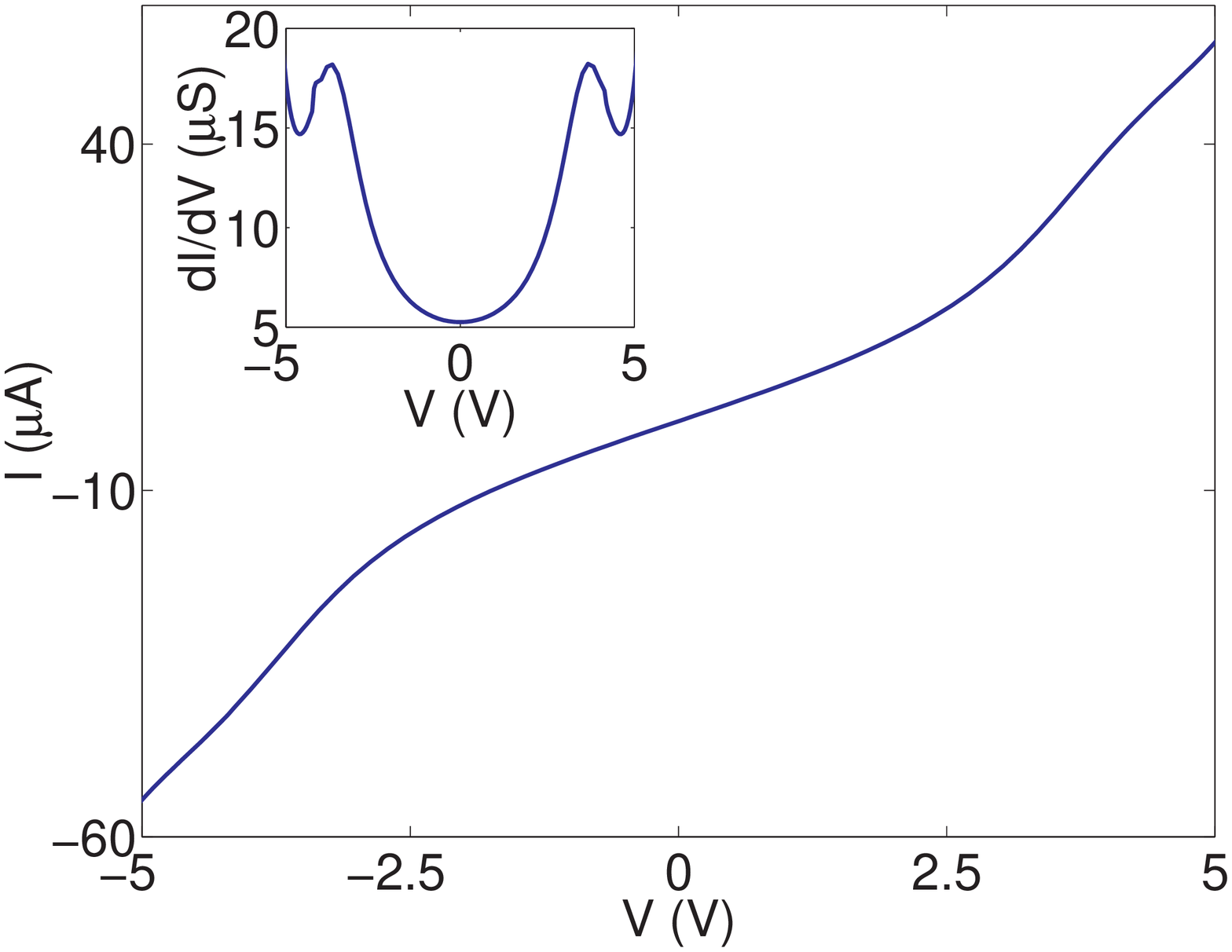,width=3.5in}}
\caption[I-V of phenyl dithiol connected to two gold contacts]{I-V of phenyl dithiol 
connected to two gold contacts. The underlying
mechanism is resonant tunneling - much like that
in a gold nanowire weakly coupled to the contacts (see Fig~\ref{iv_qpc}~b).
The gap depends on the proximity of the Fermi energy to the zero bias HOMO
level if we neglect charging effects, while the maximum current at the onset of
conduction is proportional to the parallel combination of broadenings at the
molecular energy (see text). For phenyl dithiol, the predictions from the above
estimates are about 4 volts and 40 $\mu$A respectively.}
\label{iv_pdt}
\end{figure}

{\em I-V characteristics.} A gold nanowire connected to gold contacts has a continuous 
DOS near the equilibrium Fermi energy (see Fig~\ref{cnl}, middle panel), and hence we expect it to exhibit metallic
conduction as shown in Fig~\ref{iv_qpc}~a. The nanowire 
exhibits a conductance equal
to the quantum unit conductance of $2e^2/h \approx 77 \mu S$. This ohmic I-V is a testimony to the accuracy of our
self-energy matrix. Only a correct self-energy matrix will get rid of spurious reflections at the contact and seamlessly
couple the 1-D gold wire with the 3-D gold contact. Fig~\ref{iv_qpc}~b shows the I-V characteristic 
for the same gold nanowire, but with the coupling to the contacts reduced by a
factor of 4. Due to the reduced coupling,
the broadening of the levels is much less than the separation between the levels, and the wire now
exhibits resonant conduction \cite{damghosh}. There are marked plateaus in the I-V curve when a level lies between the contact Fermi levels.
This is shown schematically in Fig~\ref{lev_cartoon}. Fig~\ref{lev_weak} shows the energy levels of the weakly coupled gold 
nanowire as a function of applied bias. At equilibrium, the contact Fermi level lies close to the LUMO level. For a 
non-zero bias $V$, $\mu_1$ and $\mu_2$ separate by an amount $eV$ 
\footnote{
We choose to split $\mu_1$ and $\mu_2$ equally around the equilibrium Fermi energy $E_f$, or $\mu_1=E_f-eV/2$ and
$\mu_2=E_f+eV/2$ (see Fig.~\ref{lev_cartoon}). This choice is made in order to be consistent with the Gaussian '98
convention of applying the linear voltage drop (through the `field' option) symmetrically across the molecule. Specifically,
for a molecule placed with it's center (along the $z$-axis) at $z=0$, Gaussian '98 applies the field such that 
one end of the molecule is at a voltage $+V/2$ and the other end is at $-V/2$. One may choose to split $\mu_1$ and $\mu_2$ 
in some other manner, and the results will not vary provided the field is applied consistently.
}
.$\mu_2$ moves closer 
to LUMO and wants to fill it up. Charging effects now come into the picture and the 
levels tend to float up and follow $\mu_2$ until $\mu_1$ comes close to crossing the HOMO level. At this point, $\mu_2$ wants 
to put charge into the LUMO while $\mu_1$ wants to empty the HOMO. The number of electrons lost is more than 
the number gained, and now the levels tend to go down with $\mu_1$ in order to prevent further loss of electrons. The current 
jumps up (see Fig~\ref{iv_qpc}~b) at that bias point where $\mu_2$ crosses the LUMO level.
Such resonant conduction is also seen in a phenyl dithiol molecule
(Fig~\ref{iv_pdt}) with pronounced peaks in the conductance $dI/dV$ at resonance.

\begin{figure}[t!]
\centerline{\psfig{figure=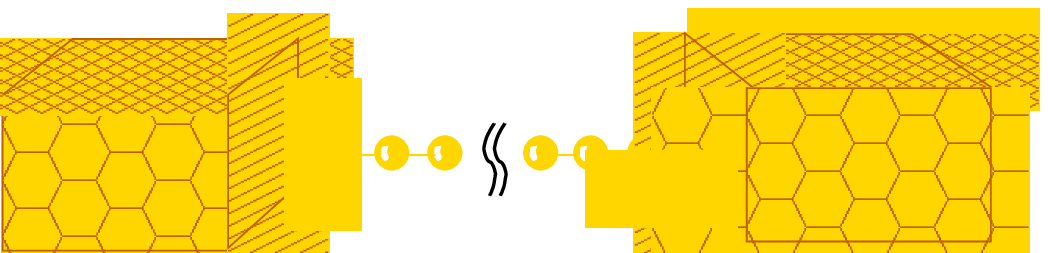,width=3.5in}}
\vspace{0.2cm}
\centerline{\psfig{figure=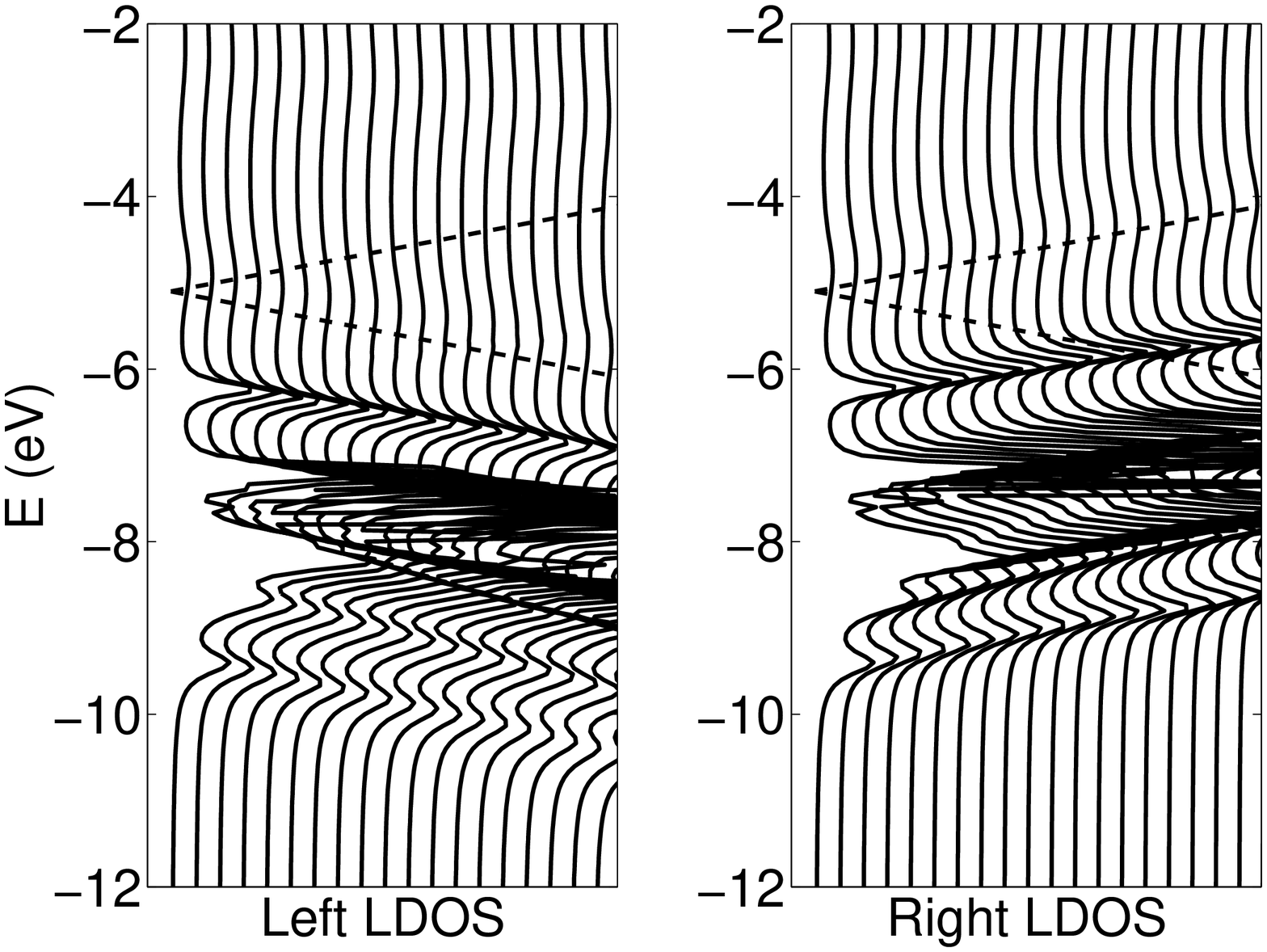,width=3.5in}}
\vspace{0.25cm}
\centerline{\psfig{figure=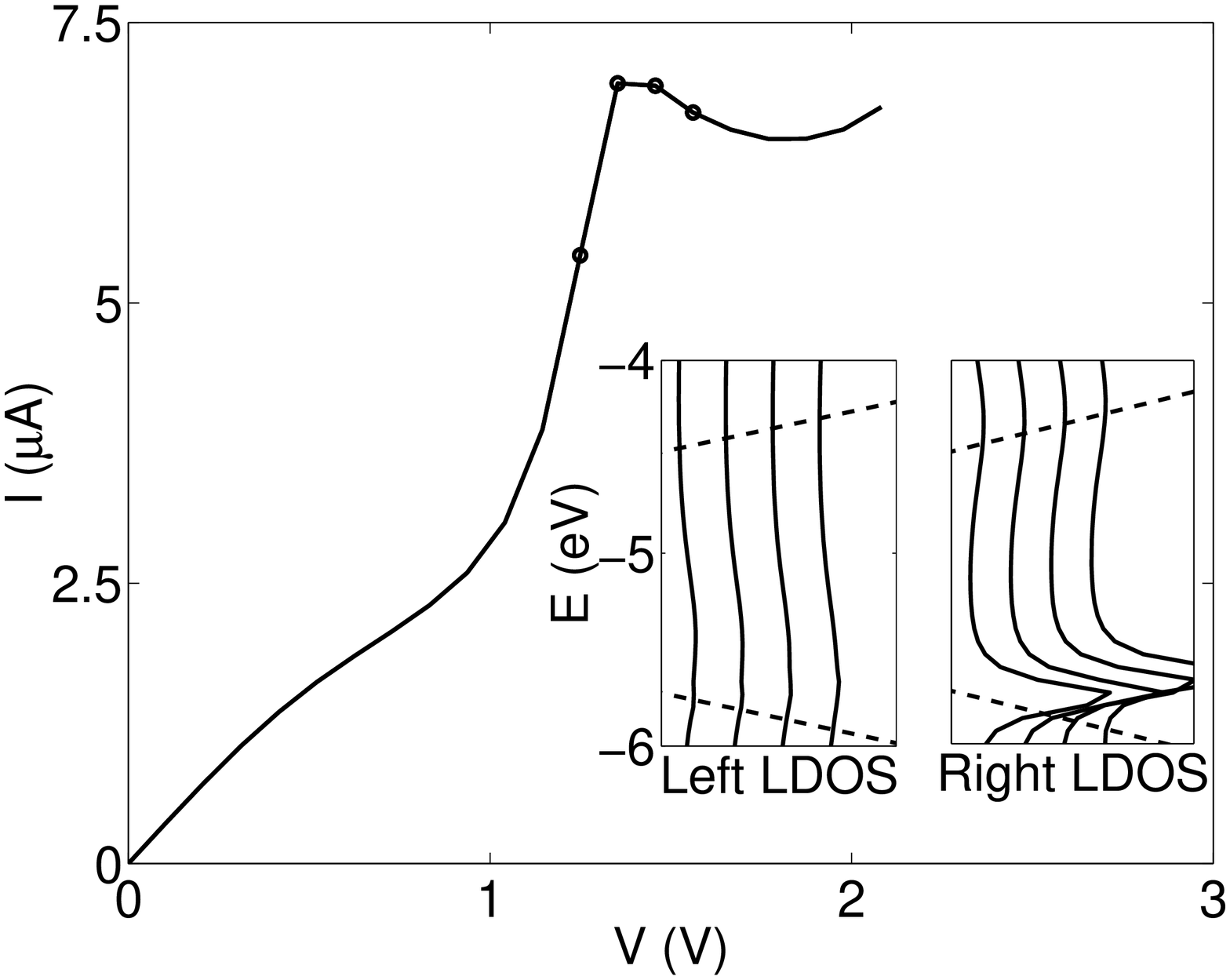,width=3.5in}}
\caption[LDOS and I-V of a gold nanowire with a stretched central bond]{Left and 
right LDOS for a wire with a defect in the center
for 21 equidistant
voltage values from 0 to 2 volts, laterally shifted equally for clarity.
The electrons in the left and right segments of the wire are separately in
equilibrium with the left and right contacts, and their respective LDOS
follow the corresponding contact electrochemical potentials (dashed lines) because
of charging effects similar to those seen
in Fig~\ref{lev_weak}.
As the two LDOS slide past each other within the $\mu_1$-$\mu_2$ window
(shown magnified in the inset of the bottom figure at the four
voltage points circled in the I-V graph) their peak values (tail of the peak for 
the left LDOS) come in and out of resonance, producing thereby a weak NDR in 
the I-V characteristic.}
\label{iv_barrier}
\end{figure}

The I-V characteristic for a quantum point contact with a defect
at the center exhibits a weak negative differential resistance (NDR) (Fig~\ref{iv_barrier}). 
We artificially stretch a bond in the middle of our
6 atom gold chain, which causes the left and right local density of
states (LDOS) of the chain to separately be in equilibrium with the
left and right contacts. Applying a bias then causes the two LDOS to
sweep past each other within $\mu_{1,2}$. Owing to the presence of some
sharp van-Hove singularities in the DOS and some smoothened out maxima,
there is a progressive alignment and misalignment of peaks in the LDOS
that leads to a weak NDR in the I-V. A similar mechanism for NDR has been observed in other calculations 
\cite{guo,magnus_paper}.

Although most semi-empirical theories qualitatively match the shape of the resonant I-V characteristics for molecular
conductors such as phenyl dithiol (PDT), quantitative agreements between experiment and theory have been largely
unsatisfactory. The I-V for PDT can be quantified by the conductance gap and the current value at the onset of conduction.
If we neglect the charging energy $U_c$ the gap is roughly given by the proximity of the contact Fermi energy to the nearest 
conducting
molecular level, $E_{gap} \approx 4|E_f - E_{mol}|$ \cite{datta_expt}; while the peak current at the onset of conduction is roughly 
given by
$2e\Gamma_1\Gamma_2/\left(\Gamma_1 + \Gamma_2\right)\hbar$, evaluated at the energy $E_{mol}$. The current onset is smeared
out over the tail of the molecular level, which in conjunction with charging effects leads to a rise in current at the
molecular level crossing extended over a voltage width $\Delta V \approx \left(\Gamma_1 + \Gamma_2 + U_C\right)/e$. Ab-initio
theories are indispensable for obtaining self-consistently the position of the Fermi energy relative to the molecular levels,
as well as the broadenings of the levels. Both from ab-initio calculations for PDT, as well as ab-initio estimates based on
the above arguments, the predicted conductance gap of PDT, set by the molecular HOMO level, comes out to about 4 Volts,
while the maximum current turns out to be tens of microamperes. Experimentally both the conductance gap and the maximum
current are much less than these theoretical estimates. The decrease in current can be attributed either to weak coupling
between molecular $\pi$ orbitals and the s-orbitals of loose gold atoms at the contact surface \cite{datta_loose,diVentra}, or 
to 
tunneling 
between
molecular units separately chemisorbed on two ends of a break-junction \cite{kirczenow_unpublished}. Thus a precise 
experimental knowledge 
of
the contact conditions is essential prior to appropriate surface modeling. The conductance gap is as yet unexplained by the
above attempts, and requires once again a precise knowledge of contact surface conditions which can vary the Fermi energy
substantially enough to alter the conductance gap.

{\em Voltage drop and electron density.} An applied
voltage is known to drop largely across the metal-molecule interface,
leading to a weaker drop in the molecule. The precise nature of the
potential profile is an important input to semi-empirical calculations
of transport. Tian et al. \cite{tian} suggested using a flat potential profile
inside the molecule, with a voltage division factor describing its
position. Such a flat profile was obtained by Mujica et al. \cite{mujica_vdrop} by solving a
1-D Poisson equation, and experimentally measured for longer ($\sim\mu$m)
wires by Seshadri and Frisbie \cite{frisbie_vdrop}. 
However,
in all these cases the geometry under consideration is a series of 2-D
charge sheets with potential variations only along the wire axis. The
1-D Poisson equation allows variations only along one coordinate, while
the measurements in \cite{frisbie_vdrop} referred to a self-assembled monolayer (SAM)
where once again transverse potential variations are screened out by the
presence of neighboring molecules. 
In contrast, Lang and Avouris \cite{lang_vdrop} obtained a significant potential drop
in a carbon atomic wire, which is a consequence of fields penetrating
from transverse directions, as correctly predicted for a break-junction
geometry by a 3-D Poisson equation.  Our particular geometry is suited
to the break-junction, since the Hartree term in Gaussian '98 is calculated
for a 3-D geometry.

\begin{figure}[h!]
\centerline{\psfig{figure=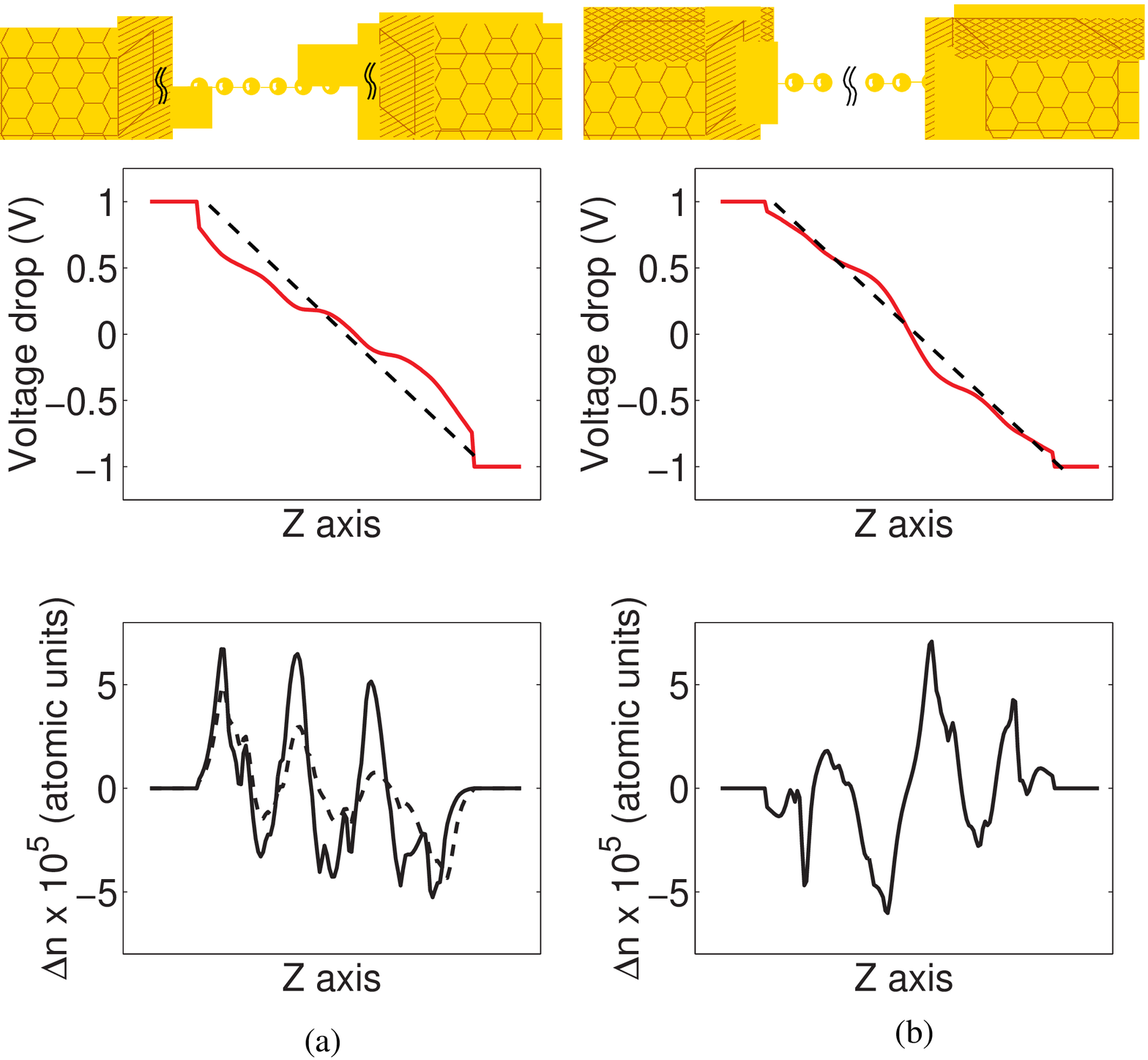,width=3.5in}}
\caption[Voltage drop in a gold nanowire coupled to gold contacts]{Electrostatics of (a) a weakly contacted gold wire, and (b)
a strongly contacted wire with a substitutional defect (a stretched bond at the 
center),
in response to a 2 volt applied bias across the contacts. Top panel:
schematic, middle panel: voltage drop along the wire (dashed lines show a linear 
drop for comparison), bottom panel:
density of electrons along the wire. In all plots, the corresponding
quantities at equilibrium have been subtracted out in order to remove
peaks near the positions of the nuclei for clarity. There is a substantial
voltage drop in the wires due to transverse fields penetrating the thin
(single atom cross-section) conductors. The largest potential drop is
at the barriers (wire-metal interfaces and defects), while the electronic
charge piles up against the applied bias as expected. Superposed on this
general polarization of charge are Friedel oscillations, which die out
(dashed line in left bottom panel) on increasing incoherent scattering in the wire,
 incorporated
through an additional phonon self-energy $\Sigma_p$ introduced through a
B\"uttiker probe.}
\label{vdrop}
\end{figure}

Fig~\ref{vdrop}~a shows a gold nanowire weakly coupled to the contacts. In Fig~\ref{vdrop}~b, the coupling to contacts is 
strong, but a 
substitutional defect has been incorporated by artificially introducing
a bond that is longer than the equilibrium gold bond length.  
The voltage
drop across the device itself is  smaller than the
applied voltage bias owing to screening effects, incorporated
self-consistently through the Hartree term of our Fock matrix. The plots on top
show the net self-consistent voltage drop across the wire
(solid lines). The electrostatic potential does not exhibit peaks at the atomic
positions because the equilibrium potential profile has been subtracted for clarity. Shown for
comparison is the linear voltage drop (dashed lines), the solution
to Laplace's equation in the region.  In (b), a large part of the
voltage drop occurs at the defect as expected. In (a) one might ask
why the voltage drops across a ballistic device.  It is important to
recognize the distinction between the electrostatic potential and the
electrochemical potential at this point. In a ballistic device, it
is the electrochemical potential that remains constant due to absence
of scattering. The electrostatic potential, however, obeys Poisson's
equation and may or may not vary depending on the charge density inside
the device. For example, a vacuum tube is a ballistic device with a linear
voltage (electrostatic potential) drop due to the absence of any charge.
The potential is not well screened out due to the thin (one atom) cross-section
of the wire, which allows transverse fields to penetrate. 
In contrast, in a self-assembled monolayer (SAM), the potential variation is expected to
resemble that predicted in \cite{tian}. However, this requires us
to disable the Hartree term in Gaussian '98 and replace it with our own 
evaluation (this is under current investigation).  

The transferred number of electrons (Fig~\ref{vdrop}, bottom panels)
$\Delta n$ along the wire obtained from our self-consistent solution
shows how the voltage drops develop. In both plots, there are
sizeable oscillations in the charge density, which we identify as
Friedel oscillations. The equilibrium electron density is $n=1/d$
for the wire ($d$ is the spacing between atoms in the wire), leading
to a 1-D Fermi wave-vector $k_F = n\pi/2 = \pi/2d$. The corresponding Friedel
oscillation wavelength is $2\pi/(2k_F)=2d$, about twice an interatomic
spacing, predicting about three oscillation cycles in a six atom
device, as we indeed observed. Furthermore, the amplitudes of the oscillations
decrease on incorporating an incoherent scattering process in the wire
through a B\"uttiker probe \cite{buttiker} (which shows up as an additional self-energy
matrix $\Sigma_p$ representing coupling of the wire with a phonon-bath,
for example).  Superposed on the Friedel oscillations is the general
trend of the charge distribution, to flow against the applied field (i.e.,
piling to the left). In (a) this piled up electronic charge screens the
applied field and reduces the slope of the voltage drop in the wire.
In (b), the charge piles up on each side of the barrier, leading to a
charge resistivity dipole in the center. This dipole has a polar field
that is now in the direction of the applied field, thus increasing the
voltage drop across the barrier .

\section{Summary}
\label{summary}

In this paper we have described a straightforward but rigorous procedure for calculating the I-V characteristics of
molecular wires. The Fock matrix is obtained using a Gaussian basis set (LANL2DZ) and the self-consistent potential is
obtained using density functional theory with the B3PW91 approximation. This approach is identical to that used in
standard quantum chemistry programs like Gaussian '98. The difference lies in our use of the NEGF formalism to calculate
the density matrix from the Fock matrix. This allows us to handle open systems far from equilibrium which are very
different from the isolated molecules commonly modeled in quantum chemistry. The close coupling with standard quantum
chemistry programs not only makes the procedure simpler to implement but also makes the relation between the I-V
characteristics and the chemistry of the molecule more obvious. Partitioning the contact-molecule-contact system into a
molecular subspace and a contact subspace allows us to focus on the quantum chemistry of the molecule while exactly taking
into account the bonding of the molecule with the contacts, contact surface physics and atomicity etc.  We use our method
to interpolate between two extreme examples of transport through a molecular wire connected to gold contacts: band
conduction in a metallic (gold) nanowire, and resonant conduction through broadened, quasidiscrete levels of a phenyl
dithiol molecule. We examine quantities of interest like the I-V characteristic, voltage drop and charge density.

We would like to thank F. Zahid, U. Savagaonkar, M. Paulsson, T. Rakshit and R. Venugopal for useful discussions. This work
was supported by NSF and the US Army Research Office (ARO) under grants number 9809520-ECS and DAAD19-99-1-0198.

\section*{Appendix~A \hspace{0.5cm} Self-energy: Imposing the FCC (111) Symmetry}

In order to calculate the gold (111) surface Green's function, we use
Gaussian '98 to simulate a 28 atom gold cluster with FCC (111) geometry 
and then extract the in-plane and out-of-plane overlap and Fock matrix 
components $S_{mn}$ and $F_{mn}$ (see the discussion following 
Eq.~\ref{FakSak} in Section~\ref{theory}). The cluster, however, does not have
the full group theoretical symmetry of the FCC (111) crystal due to edge-effects.
This symmetry needs to be imposed on the extracted Fock matrix elements in order to be 
consistent with the assumptions of two-dimensional periodicity that go into our
$\vec{k}$-space formalism. The overlap matrix elements automatically satisfy these
symmetry requirements, since they involve products of two on-site localized orbitals.

A typical cluster is shown in Fig~\ref{cluster}, with atoms numbered from 1 to 13.
An FCC (111) surface consists of repetitions of such clusters in the sequence 
ABCABCABC $\ldots$, B being the plane numbered 1 to 7, and A and C being the
planes with the rotated triangles above and below it. Each plane is parallel to the
x-y plane, atom 1 is at the origin and atom 2 lies on the x-axis. This gives us a
$z$-axis in the (111) direction, along which the molecule is assumed to align after
contacting with the Au surface. The Fock matrix for the cluster looks as follows:

\begin{equation}
F = \left[\begin{array}{cccc} F_{11} & F_{12} & \cdots&F_{1,13}\\
F_{21}&F_{22} &\ldots \\
\ldots\\
\ldots\\
\ldots\end{array}\right]
\label{eVI1}
\end{equation}
The LANL2DZ basis for gold consists of three sets of s, three sets of p and 2 sets of
d orbitals, giving us $3 + 3 \times 3 + 2 \times 5 = 22$ orbitals on each gold atom. 
Each of the above matrix components $F_{11}$, $F_{12}$ etc. is therefore 22 $\times$
22 in size. Using Gaussian '98, we can extract all these matrix components separately.
For a Au(111) surface, however, the elements should be interrelated by the point group
symmetries of the FCC crystal. Starting with two independent matrix components, say,
an on-site element $F_{11}$ that is characteristic of the gold atom, and a nearest
neighbor element $F_{12}$ that depends on the lattice parameter, we construct the
rest of the elements of $F$ using the symmetry operations. 

The FCC crystal has the following symmetry: on rotating the cluster in 
Fig.~\ref{cluster} by 60 degrees about the $z$-axis followed by a reflection in 
the x-y plane, we get back the same cluster. This means there is a unitary matrix
$M$ that takes $F_{12}$ into $F_{13}$ and so on. The matrix $M$ can be block-decomposed
into contributions from each of the s, p and d orbital subspaces. Since s orbitals are
spherically symmetric, the corresponding block for the 3 s orbitals is the 3 $\times$
3 identity matrix. For p orbitals, the $M_p$ matrix consists of the following 
reflection (R) and rotation (U) matrices:

\begin{eqnarray}
M_p &=& R_pU_p \nonumber\\
R_p &=& \left(\begin{array}{ccccc} 1 &  & 0 & & 0\\ 0 & & 1 && 0 \\ 0 && 0 && -1\end{array}\right)\nonumber\\
U_p &=& \left(\begin{array}{ccccc} \cos{\pi/3} & & -\sin{\pi/3} && 0 \\ \sin{\pi/3}
&& \cos{\pi/3} && 0 \\ 0 && 0 && 1\end{array}\right)  
\label{ep}
\end{eqnarray}
A similar matrix can be constructed for the d-orbital subspace. At the end of the
process we thus get a 22 $\times$ 22 $M$ matrix describing rotation-reflection of 
the entire LANL2DZ basis set.

The nearest neighbor in-plane couplings of atom 1 can now be generated by repeated
applications of the rotation-reflection transformation:
\begin{eqnarray}
F_{17} &=& M^\dagger F_{12}M\nonumber\\
F_{16} &=& M^\dagger F_{17}M\nonumber\\
F_{15} &=& M^\dagger F_{16}M\nonumber\\
F_{14} &=& M^\dagger F_{15}M\nonumber\\ 
F_{13} &=& M^\dagger F_{14}M\nonumber\\
F_{12} &=& M^\dagger F_{13}M
\label{eVI2}
\end{eqnarray}
A simple out-of-plane rotation by 120 degrees can generate $F_{18}$ from $F_{12}$
\footnote{
For the results shown in this paper, we used the $F_{18}$ obtained from Gaussian '98 instead of generating it from 
$F_{12}$. We thus have three independent matrix elements ($F_{11}$, $F_{12}$ and $F_{18}$) instead of two ($F_{11}$ and $F_{12}$).
$F_{18}$ has the same structure as that of $F_{12}$ (see Eq.~\ref{eP}). Preliminary calculations with $F_{18}$ generated from 
$F_{12}$ using Eq.~\ref{eM} do not show a significant change in the results.}
. Once again, we explicitly
write down the contribution for the p orbitals. The d orbitals need to be considered analogously while the s
orbitals remain unchanged.
\begin{eqnarray}
F_{18} &=& (M^\prime)^\dagger F_{12} M^\prime \nonumber\\
M^\prime_p &=& \left(\begin{array}{ccccc} -1/2 && -1/2\sqrt{3} && \sqrt{2/3} \\ 1/2\sqrt{3} && 5/6 && \sqrt{2}/3
\\ -\sqrt{2/3} && \sqrt{2}/3 && -1/3\end{array}\right)
\label{eM}
\end{eqnarray}
All the other matrix components can be similarly obtained, using translation and rotation-reflection invariance.
For example:
\begin{eqnarray}
F_{1,12} &=& M^\dagger F_{18} M \nonumber\\
F_{1,10} &=& M^\dagger F_{1,12} M \nonumber\\
F_{22} &=& F_{11} \nonumber \\
F_{23} &=& F_{14} 
\label{eN}
\end{eqnarray}
and so on.

The crystalline symmetries also constrain the structures of the two fundamental matrix components $F_{11}$ and 
$F_{12}$ out of which all other components are constructed. $F_{11}$ is a diagonal block of $F$ and is Hermitian.
$F_{12}$ satisfies the following condition:

\begin{equation}
\left(M^\dagger\right)^3F_{12}M^3 = F_{12}^\dagger
\label{eO}
\end{equation}
Three operations of the rotation-reflection transformation (equivalently, a coordinate inversion) on $F_{12}$ 
generates the matrix $F_{15}$, which equals $F_{51}^\dagger$ by bond-inversion symmetry. Translational symmetry
implies $F_{51} = F_{12}$, which leads to the above equation. The form of $F_{12}$ consistent with the above
condition is given by:

\begin{equation}
F_{12} = \left( \begin{array} {ccc} a & b & c \\ -b^\dagger & e & d \\ c^\dagger & -d^\dagger & f \end{array}\right)
\label{eP}
\end{equation}
where $a$, $e$ and $f$ are hermitian matrices. The $F_{12}$ matrix extracted from Gaussian will have this structure
for a large enough cluster, else we need to impose this condition as well. 

With a properly symmetrized set $F_{mn}$ and $S_{mn}$, the matrices $F_{a\vec{k}}$ and $S_{a\vec{k}}$ in 
Eq.~\ref{FakSak} are Hermitian as expected, and we may proceed to calculate the gold (111) surface Green's 
function as outlined in Section~\ref{theory}. Ignoring the symmetries, however, can lead to non-physical results,
including negative density of states or unsymmetric Green's function matrices.

\section*{Appendix~B \hspace{0.5cm} Density Matrix: Analytical Integration}                     

We have developed a fast and elegant algorithm to evaluate the density matrix given an energy independent self-energy. For
bulk gold it is known that around the Fermi energy the local DOS is approximately a constant
\cite{papaconst}. It is reasonable to expect that all the transport properties of the molecule are determined by the molecular
levels near $E_f$ and the deep levels contribute to the total number of electrons by remaining full in the bias range of
interest. In view of this, We split the energy range of integration in two parts and denote the boundary by $E_b$. $E_b$ is
chosen such that it is a few electron-volts below the Fermi energy $E_f$ and the molecular density of states at $E_b$ is
negligible. We replace $\Sigma(E)$ by $\Sigma(E_f)$ in the energy range between $E_b$ to $E_f$. 
\footnote{If the DOS is finite at $E_b$, there will be a dependence of our results
on $E_b$, leading to a corresponding ambiguity regarding the precise location
of $E_f$ at equilibrium. This is an issue that deserves more attention,
since the precise location of $E_f$ is altered significantly even if the total
number of electrons, integrated over the entire energy range, is miscalculated
by a small fraction. As such, the constant self-energy approximation used
here could also influence $E_f$ in a similar way.}
For the energy range from
$-\infty$ to $E_b$ we replace $\Sigma(E)$ by $-i\eta$, a constant infinitesimal broadening. The only energy dependence in the
integrand in Eq.~\ref{rho} now appears through the Green's function $G$ and we can perform an analytical integration in an
appropriate eigenspace, as explained below.

To evaluate the density matrix, we now need to solve integrals of the type (see equations \ref{rho} through \ref{e11} in 
Section~\ref{theory})
\begin{equation}
2\pi \rho = \int _{-\infty}^{\infty} dE~f(E)~G(E)~\Gamma ~G^\dagger (E)
\label{appA1}
\end{equation}
where $\Gamma$ is the anti-hermitian part (see Eq.~\ref{e11}) of the energy independent self-energy and $f(E)$ is the Fermi
function with some electrochemical potential $\mu$ (see Eq.~\ref{e9}). In this paper we assume $T=0~K$ which modifies the
integral in the above equation as
\begin{equation}
2\pi \rho = \int _{E_{min}}^{\mu} dE~G(E)~\Gamma ~G^\dagger (E)
\label{appA2}
\end{equation}
where it is understood that $E_{min}$ tends to $-\infty$. Using Eq.~\ref{e10} for $G(E)$ we may rewrite the above 
equation as
\begin{equation}
2 \pi \rho = S^{-\frac{1}{2}} \left [ \int _{E_{min}}^{\mu} dE~\overline{G} (E)~\overline{\Gamma}~\overline{G} ^\dagger (E) \right ] S^{-\frac{1}{2}}
\label{appA3}
\end{equation} 
where $S$ is the overlap matrix $\overline{\Gamma}$ and $\overline{G} (E)$ are defined as
\[
\overline{\Gamma} = S^{-\frac{1}{2}} \Gamma S^{-\frac{1}{2}}
\]
and 
\[
\overline{G} (E) = \left( EI - \overline{F} \right ) ^{-1}
\]
with $I$ being the identity matrix and 
\[
\overline{F} = S^{-\frac{1}{2}} \left ( F + \Sigma _1 + \Sigma _2 \right ) S^{-\frac{1}{2}} 
\]
In the above equation, $\Sigma _{1,2}$ are the energy independent self-energy matrices as discussed at the beginning of this 
appendix and $F$ is the molecular Fock matrix.

To solve the integral in Eq.~\ref{appA3} analytically, we work in the eigenspace of the energy independent non-Hermitian 
matrix $\overline{F}$ given by the above equation. Since $\overline{F}$ is non-Hermitian, we need its eigenkets $\{|m\rangle\}$
and their dual kets $\{|\tilde{m} \rangle\}$ in order to form a complete basis set. The corresponding eigenvalues are then 
$\{\epsilon_m\}$ and $\{\epsilon^*_m\}$. Expanding in this mixed basis set, we get
\begin{eqnarray}
\noindent & 2\pi \rho&   \nonumber\\
&=& S^{-\frac{1}{2}} \int _{E_{min}} ^{\mu} dE ~
\overline{G}(E)\left(\sum_p|p\rangle \langle\tilde{p}|\right) \nonumber\\
&\times& \overline{\Gamma}~
\overline{G}^\dagger (E)\left(\sum_q|\tilde{q} \rangle \langle q|\right)
S^{-\frac{1}{2}} \nonumber\\
&=& S^{-\frac{1}{2}} \left[ \sum_{pq} \overline{\Gamma}_{\tilde{p}\tilde{q}} |p\rangle
\langle q| \int _{E_{min}} ^{\mu} {{dE}\over{\left(E - \epsilon_p\right)
\left(E - \epsilon^*_q\right)}} \right ] S^{-\frac{1}{2}}\nonumber\\
\label{appA4}
\end{eqnarray}
where we used $\overline{G} (E)|m\rangle = |m\rangle/(E-\epsilon_m)$ and $\overline{G} ^\dagger (E) |\tilde{m} \rangle= |\tilde{m} 
\rangle /(E-\epsilon^*_m)$. 

It is now straightforward to do the above integral analytically and obtain the density matrix. The current $I$ given by
Eq.~\ref{e13} may also be evaluated analytically using a similar procedure. This analytical procedure allows us to do the 
integration extremely fast and avoids errors arising out of a numerical integration over a grid.

\bibliography{main}

\begin{thebibliography}{10}
\expandafter\ifx\csname url\endcsname\relax
  \def\url#1{\texttt{#1}}\fi
\expandafter\ifx\csname urlprefix\endcsname\relax\def\urlprefix{URL }\fi

\bibitem{reed_review}
{For a review of the experimental work see M.A. Reed}, Proc. IEEE 87 (1999)
  652.

\bibitem{eigler}
A.~Yazdani, D.~Eigler, N.~Lang, Science 272 (1996) 1921.

\bibitem{collier}
C.~Collier, E.~Wong, M.~Belohradsky, F.~Raymo, J.~Stoddart, P.~Kuekes,
  R.~Williams, J.~Heath, Science 285 (1999) 391.

\bibitem{chen}
J.~Chen, M.~Reed, A.~Rawlett, J.~Tour, Science 286 (1999) 1550.

\bibitem{reed_expt}
M.~Reed, C.~Zhou, C.~Muller, T.~Burgin, J.~Tour, Science 278 (1997) 252.

\bibitem{zhou_reed_expt}
C.~Zhou, M.~Deshpande, M.~Reed, L.~Jones, J.~Tour, Appl.Phys.Lett. 71 (1997)
  2857.

\bibitem{andres_expt}
R.~Andres, T.~Bein, M.~Dorogi, S.~Feng, J.~Henderson, C.~Kubiak, W.~Mahoney,
  R.~Osifchin, R.~G. Reifenberger, Science 272 (1996) 1323.

\bibitem{datta_expt}
S.~Datta, W.~Tian, S.~Hong, R.~Reifenberger, J.~I. Henderson, C.~Kubiak, Phys.
  Rev. Lett. 79 (1997) 2530.

\bibitem{durig}
U.~Durig, O.~Zuger, B.~Michel, L.~Haussling, H.~Ringsdorf, Phys. Rev. B 48
  (1993) 1711.

\bibitem{boulas}
C.~Boulas, J.~Davidovits, F.~Rondelez, D.~Vuillaume, Phys. Rev. Lett. 76 (1996)
  4797.

\bibitem{ohnishi}
H.~Ohnishi, Y.~Kondo, K.~Takayanagi, Nature 395 (1998) 780.

\bibitem{yanson}
A.~Yanson, G.~Bollinger, H.~van~den Brom, N.~Agrait, J.~van Ruitenbeek, Nature
  395 (1998) 783.

\bibitem{dekker_phys_today}
C.~Dekker, Phys. Today 52 (1999) 22.

\bibitem{fink_dna}
H.~Fink, C.~Schonenberger, Nature 398 (1999) 407.

\bibitem{porath_dna}
D.~Porath, A.~Bezryadin, S.~de~Vries, C.~Dekker, Nature 403 (2000) 635.

\bibitem{ratner_dna}
M.~Ratner, Nature 397 (1999) 480.

\bibitem{schon_samfet}
J.~Sch\"on, H.~Meng, Z.~Bao, Nature 413 (2001) 713.

\bibitem{tian}
W.~Tian, S.~Datta, S.~Hong, R.~Reifenberger, J.~I. Henderson, C.~Kubiak, J.
  Chem. Phys. 109 (1997) 2874.

\bibitem{magnus_paper}
M.~Paulsson, S.~Stafstr\"om, Phys. Rev. B 64 (2001) 035416.

\bibitem{emberly_prb_58}
E.~Emberly, G.~Kirczenow, Phys. Rev. B 58 (1998) 10911.

\bibitem{hush}
L.~E. Hall, J.~R. Reimers, N.~S. Hush, K.~Silverbrook, J. Chem. Phys. 112
  (2000) 1510.

\bibitem{yaliraki}
S.~Yaliraki, A.~Roitberg, C.~Gonzalez, V.~Mujica, M.~Ratner, J. Chem. Phys. 111
  (1999) 6997.

\bibitem{mujica_vdrop}
V.~Mujica, A.~Roitberg, M.~Ratner, J. Chem. Phys. 112 (2000) 6834.

\bibitem{tsu_theory}
C.~Tsu, R.~Marcus, J. Chem. Phys. 106 (1997) 584.

\bibitem{magoga_theory}
M.~Magoga, C.~Joachim, Phys.Rev. B 56 (1997) 4722.

\bibitem{onipko_theory}
A.~Onipko, Phys. Rev. B 59 (1999) 9995.

\bibitem{diVentra}
{M. Di Ventra, S.T. Pantelides and N.D. Lang }, Phys. Rev. Lett. 84 (2000) 979.

\bibitem{lang_vdrop}
N.~Lang, P.~Avouris, Phys. Rev. Lett. 84 (2000) 358.

\bibitem{guo}
B.~Larade, J.~Taylor, H.~Mehrez, H.~Guo, Phys. Rev. B 64 (2001) 075420.

\bibitem{seminario}
J.~Seminario, A.~Zacarias, J.~Tour, J.Phys.Chem. A 103 (1999) 7883.

\bibitem{palacios}
{J. J. Palacios, A. J. P\'erez-Jim\'enez, E. Louis and J. A. Verg\'es,
  cond-mat/0101359}.

\bibitem{szabo}
A.~Szabo, N.~Ostlund, Modern Quantum Chemistry, Dover Publications, Inc.,
  Mineola, New York, 1996.

\bibitem{jensen}
F.~Jensen, Introduction to Computational Chemistry, Wiley, 1999.

\bibitem{datta_book}
S.~Datta, Electronic Transport in Mesoscopic Systems, Cambridge University
  Press, 1995.

\bibitem{datta_tut}
S.~Datta, Superlattices and Microstructures 28 (2000) 253.

\bibitem{gaussian}
{Gaussian 98 (Revision A.7), M. J. Frisch, G. W. Trucks, H. B. Schlegel, G. E.
  Scuseria, M. A. Robb, J. R. Cheeseman, V. G. Zakrzewski, J. A. Montgomery,
  Jr., R. E. Stratmann, J. C. Burant, S. Dapprich, J. M. Millam, A. D. Daniels,
  K. N. Kudin, M. C. Strain, O. Farkas, J. Tomasi, V. Barone, M. Cossi, R.
  Cammi, B. Mennucci, C. Pomelli, C. Adamo, S. Clifford, J. Ochterski, G. A.
  Petersson, P. Y. Ayala, Q. Cui, K. Morokuma, D. K. Malick, A. D. Rabuck, K.
  Raghavachari, J. B. Foresman, J. Cioslowski, J. V. Ortiz, A. G. Baboul, B. B.
  Stefanov, G. Liu, A. Liashenko, P. Piskorz, I. Komaromi, R. Gomperts, R. L.
  Martin, D. J. Fox, T. Keith, M. A. Al-Laham, C. Y. Peng, A. Nanayakkara, C.
  Gonzalez, M. Challacombe, P. M. W. Gill, B. G. Johnson, W. Chen, M. W. Wong,
  J. L. Andres, M. Head-Gordon, E. S. Replogle and J. A. Pople, Gaussian, Inc.,
  Pittsburgh PA} (1998).

\bibitem{lanl2dz_1}
P.~Hay, W.~Wadt, J. Chem. Phys. 82 (1985) 270.

\bibitem{lanl2dz_2}
P.~Hay, W.~Wadt, J. Chem. Phys. 82 (1985) 284.

\bibitem{becke}
A.~D. Becke, J. Chem. Phys. 98 (1993) 5648.

\bibitem{perdew}
J.~P. Perdew, in: P.~Ziesche, H.~Eschrig (Eds.), Electronic structure of
  solids, Akademic Verlag, Berlin, 1991, pp. 11--20.

\bibitem{damghosh}
P.~S. Damle, A.~W. Ghosh, S.~Datta, Phys. Rev. B (Rapid Comms.) 64 (2001)
  201403.

\bibitem{xue_paper}
Y.~Xue, S.~Datta, M.~Ratner, J. Chem. Phys. 115 (2001) 4292.

\bibitem{larsen}
N.~B. Larsen, H.~Biebuyck, E.~Delamarche, B.~Michel, J. Am. Chem. Soc. 119
  (1997) 3107.

\bibitem{camillone}
N.~Camillone, C.~E.~D. Chidsey, G.~Y. Liu, G.~Scoles, J. Phys. Chem. 98 (1993)
  3503.

\bibitem{xue_thesis}
Y.~Xue, Ph.D. thesis, Purdue University (2000).

\bibitem{manoj_thesis}
M.~Samanta, Master's thesis, Purdue University (1995).

\bibitem{yang_frac_el}
W.~Yang, Y.~Zhang, Phys. Rev. Lett. 84 (2000) 5172.

\bibitem{russier_frac_el}
V.~Russier, Phys. Rev. B 45 (1992) 8894.

\bibitem{lowdin}
{See pages 152, 203 in \cite{szabo} for a discussion on Mulliken/L\"owdin
  partitioning}.

\bibitem{contour}
R.~Zeller, J.~Deutz, P.~Dederichs, Solid State Comm. 44 (1982) 993.

\bibitem{buttiker}
M.~B\"uttiker, IBM J. Res. Dev. 32 (1988) 63.

\bibitem{li_dna}
X.-Q. Li, Y.~Yan, Appl. Phys. Lett. 79 (2001) 2190.

\bibitem{tersoff}
J.~Tersoff, Phys. Rev. B 30 (1984) 4874.

\bibitem{datta_loose}
S.~Datta, D.~Janes, R.~Andres, C.~Kubiak, R.~Reifenberger, Semicond. Sci. Tech.
  13 (1998) 1347.

\bibitem{kirczenow_unpublished}
{E.G. Emberly and G. Kirczenow, unpublished} (2001).

\bibitem{frisbie_vdrop}
K.~Seshadri, C.~Frisbie, Appl. Phys. Lett. 78 (2001) 993.

\bibitem{papaconst}
D.~Papaconstantopoulos, Handbook of the Band Structure of Elemental Solids,
  Plenum Press, New York, 1986.

\end{thebibliography}
\newpage
\newpage

\end{document}